\documentclass[12pt,oneside,a4paper]{article}
\usepackage{alpha}
\usepackage{placeins}
\usepackage{float}
\usepackage{macros_alpha}
\usepackage{changepage}  
\usepackage{pgf} 
     
\usebiblio{paper} 
\begin{document}
\preprintno{
%IJCLab/20-XX\\
%LPSC/20-XX\\
MS-TP-21-28
}

\title{%
{Extraction of $B_s \to D^{(*)}_s$ form factors from ${ N_f}=2$ lattice QCD}
}

\author[Orsay]{Beno\^it~Blossier}
\author[Grenoble]{Pierre~Henri~Cahue}
\author[wwu]{Jochen~Heitger} 
\author[Orsay]{Simone~La~Cesa}
\author[Orsay,wwu]{Jan~Neuendorf}
\author[CPT]{Savvas Zafeiropoulos}

\address[Orsay]{Laboratoire de Physique des 2 Infinis Ir\`ene Joliot-Curie, CNRS/IN2P3, Universit\'e Paris-Saclay,  B\^atiment 210, 91405 Orsay Cedex, France}
\address[Grenoble]{Laboratoire de Physique Subatomique et de Cosmologie (LPSC), Universit\'e Grenoble-Alpes, CNRS/IN2P3, 53 Avenue des Martyrs, F-38026 Grenoble, France}
\address[wwu]{Institut~f\"ur~Theoretische~Physik, Westf\"alische Wilhelms-Universit\"at M\"unster, Wilhelm-Klemm-Str.~9, 48149~M\"unster, Germany}
\address[CPT]{Aix Marseille Univ, Universit\'e de Toulon, CNRS, CPT, Marseille, France}

\begin{abstract}
We report on a two-flavour lattice QCD study of the $B_s \to D_s$ and $B_s \to D^*_s$ transitions parameterized, in the heavy quark limit, by
the form factors ${\cal G}$, and $h_{A_1}$, $h_{A_2}$ and $h_{A_3}$, respectively. In the search for New Physics 
through tests of lepton flavour universality,
$B_s$ decay channels are complementary to the $B$ decays widely studied at $B$ factories and LHCb, while on the theory side 
they can be better controlled than 
the $B_c$ and $\Lambda_b$ decays. The purpose of this exploratory two-flavour study is, 
in preparation for future analyses of lattice QCD simulations with ${N_f>2}$ and physical quark-masses, 
to gain experience on a suitable method for a lattice extraction of form factors associated with $b \to c$ currents that may yield
tighter control over systematic effects like contamination from excited states and cut-off effects. We obtain the zero-recoil values 
${\cal G}^{B_s \to D_s}(1)=1.03(14)$ and $h^{B_s \to D^*_s}(1)=0.85(16)$.
\end{abstract}

%\begin{keyword}
%Lattice QCD \sep Heavy Quark Effective Theory \sep b-quark mass
%\PACS{% 
%12.38.Gc\sep %Lattice QCD calculations
%12.39.Hg\sep %Heavy quark effective theory
%14.65.Fy\sep %Bottom quarks
%12.15.Ff}    %masses and mixing (electroweak interactions)
%\end{keyword}

\maketitle
%%%%%%%%%%%%%%%%%%%%%%%%%%%%%%%%%%%%%%%
\section{\label{Introduction}Introduction}
%%%%%%%%%%%%%%%%%%%%%%%%%%%%%%%%%%%%%%%

\def\paperorletter{letter}
Though the Standard Model of particle physics (SM) has shown to describe the fundamental interactions up to the electroweak scale pretty well, experimental measurements sometimes give 
results that were totally unexpected. In particular, a recent and spectacular example in flavour physics phenomenology is the so-called $B$ anomalies in the
test of lepton flavour universality in $b \to c$ semileptonic decays. The ratio of decay widths $R_{H_c} \equiv \frac{\Gamma(H_b \to H_c \tau \nu_\tau)}{\Gamma(H_b \to H_c \ell \nu_\ell)_{\ell = e,\mu}}$ has been considered in cases $(H_b, H_c) = (B, D)$, $(B, D^*)$ and $(B_c, J/\psi)$ and compared with theoretical expectations. A discrepancy of  $2\sigma$ has been observed for $R_D$ \cite{Lees:2012xj, Lees:2013uzd, Huschle:2015rga, Aaij:2015yra, Sato:2016svk, Lattice:2015rga, Na:2015kha, Bigi:2016mdz, 
Bernlochner:2017jka, Jaiswal:2017rve}, $3\sigma$ for $R_{D^*}$ \cite{Hirose:2016wfn, Hirose:2017dxl, Aaij:2017uff, Aaij:2017deq, Bigi:2016mdz, Bernlochner:2017jka, Bigi:2017jbd, 
Jaiswal:2017rve} and  $\sim \sigma$ for $R_{J/\psi}$ \cite{Aaij:2017tyk, Harrison:2020nrv}. A bunch of New Physics models have been advocated to explain those
discrepancies, together with other anomalies in $b \to s \ell \ell$ transitions, like models with extra doublets of Higgs fields \cite{Fajfer:2012vx} or with leptoquarks 
\cite{DiLuzio:2017vat}. Further measurements are led to explore more $b \to c$ processes, such as $\Lambda_b \to \Lambda_c l \nu_l$ or $B_s \to D^{(*)}_s l \nu$, and are
expected to come soon at LHCb and Belle 2. 

A theoretical effort has been undertaken to extract the form factors associated with $B_s \to D^{(*)}_s l \nu$, but the number of
results is still limited \cite{Atoui:2013mqa, Monahan:2017uby, Harrison:2017fmw, McLean:2019sds, McLean:2019qcx, Harrison:2021tol}.
Our aim here is to study whether Wilson-Clover fermions, in combination with the step-scaling in mass method \cite{Blossier:2009hg, Atoui:2013mqa}, provide a suitable
lattice regularisation to get reliable results for these processes, as far as cut-off effects and contamination by excited 
states are concerned. While our current exploratory work focuses on   $N_f=2$, 
any phenomenologically relevant application of the methods investigated here 
has to be done in lattice QCD with $N_f=2+1(+1)$ flavours.

The paper is organised as follows: in Section \ref{sec2} we highlight the strategy we have come up with, in Section \ref{sec3}
we give the simulation details and collect our raw data. In Section \ref{sec4} we present our analysis and comment on the results. 
Section \ref{sec5} contains the conclusion.

%%%%%%%%%%%%%%%%%%%%%%%%%%%%%%%%%%%%%%%
\section{\label{sec2}Strategy}
%%%%%%%%%%%%%%%%%%%%%%%%%%%%%%%%%%%%%%%

\def\paperorletter{letter}

The starting point is to consider the hadronic matrix element $\langle D_s(p_{D_s}) | V^\mu|H_s(p_{H_s})\rangle$, with $V^\mu = \bar{c}\gamma^\mu h$. Its Lorentz structure decomposition reads
\begin{equation}
\langle D_s(p_{D_s}) | V^\mu|H_s(p_{H_s})\rangle=f^+(q^2) (p_{D_s}+p_{H_s})^\mu + f^-(q^2)q^\mu,
\end{equation}
with the quantities $f^+$ and $f^-$ familiar as form factors depending on
$q=p_{H_s}-p_{D_s}$. We will be interested in the physics case $H_s \equiv B_s$. In phenomenology it is convenient to introduce the scalar form factor $f_0$ defined by:
\begin{equation}
\langle D_s(p_{D_s}) | V^\mu|H_s(p_{H_s})\rangle=f^+(q^2) \left[(p_{D_s}+p_{H_s})^\mu - \frac{m^2_{H_s}-m^2_{D_s}}{q^2} q^\mu\right]
+ f_0(q^2)\frac{m^2_{H_s}-m^2_{D_s}}{q^2} q^\mu,
\end{equation}
Hence, we have 
\begin{eqnarray}
f^-(q^2)&=&\frac{m^2_{H_s}-m^2_{D_s}}{q^2}[f_0(q^2)-f^+(q^2)],\\
f_0(q^2)&=&f^+(q^2)+f^-(q^2)\frac{q^2}{m^2_{H_s}-m^2_{D_s}}.
\end{eqnarray}
With the kinematical configuration of an $H_s$-meson at rest, we can write
\begin{eqnarray}
\nonumber
\langle D_s(p_{D_s}) | V^0|H_s(p_{H_s})\rangle&=&f^+(q^2) (E_{D_s}+m_{H_s}) + f^-(q^2)(m_{H_s}-E_{D_s}),\\
\langle D_s(p_{D_s}) | V^i|H_s(p_{H_s})\rangle&=&-q^i (f^+(q^2)-f^-(q^2)).
\end{eqnarray}
When the $H_s$ meson is put at rest, 
the recoil $w=\frac{p_{D_s}\cdot p_{H_s}}{m_{H_s}m_{D_s}}\equiv v_{H_s} \cdot v_{D_s}$ has the simple expression $w=E_{D_s}/m_{D_s}$. Our goal is
to extract the form factor $f^+(B_s \to D_s)$ at zero recoil, $w=1$. However, except in the elastic case $D_s \to D_s$ where one finds trivially that $f^+(w=1)=1$, this kinematical point is not directly accessible to a lattice measurement. It is only by extrapolating in $w$ that one can obtain results at zero recoil. With $\vec{q}=-(\theta,\theta,\theta)$, $w\sim \sqrt{m_{D_s}^2+3\theta^2}/m_{D_s}$, we have
\begin{eqnarray}
\nonumber
f^+(w)&=&\frac{1}{2m_{H_s}}\left[\langle D_s(p_{D_s}) | V^0|H_s(p_{H_s})\rangle +
\frac{m_{H_s}-E_{D_s}}{3\theta}\sum_i \langle D_s(p_{D_s}) | V^i|H_s(p_{H_s})\rangle\right],\\
\nonumber
f^-(w)&=&\frac{1}{2m_{H_s}}\left[\langle D_s(p_{D_s}) | V^0|H_s(p_{H_s})\rangle -
\frac{m_{H_s}+E_{D_s}}{3\theta} \sum_i \langle D_s(p_{D_s}) | V^i|H_s(p_{H_s})\rangle\right].\\
\end{eqnarray}
On the other hand, in the framework of Heavy Quark Effective Theory (HQET), it is also common to parameterise the matrix element in question in terms of the form factors $h^\pm(w)$,
\begin{equation}
\frac{\langle D_s(p_{D_s}) | V^\mu|H_s(p_{H_s})\rangle}{\sqrt{m_{D_s}m_{H_s}}}=h^+(w) (v_{D_s}+v_{H_s})^\mu + h^-(w)(v_{H_s}-v_{D_s})^\mu,
\end{equation}
\begin{eqnarray}
\nonumber
h^+(w)&=&\frac{1}{2\sqrt{m_{D_s}m_{H_s}}} \left[(m_{H_s}+m_{D_s}) f^+(w)+(m_{H_s}-m_{D_s}) f^-(w)\right],\\
h^-(w)&=&\frac{1}{2\sqrt{m_{D_s}m_{H_s}}} \left[(m_{H_s}-m_{D_s}) f^+(w)+(m_{H_s}+m_{D_s}) f^-(w)\right],
\end{eqnarray}
with the aid of which one introduces the factors ${\cal G}(w)$ and $H(w)$:
\begin{eqnarray}
\nonumber
{\cal G}(w)&=&h^+(w)\left[1-\frac{m_{H_s}-m_{D_s}}{m_{H_s}+m_{D_s}} \frac{h^-(w)}{h^+(w)}\right]\\
&=&h^+(w)\left[1-\left(\frac{m_{H_s}-m_{D_s}}{m_{H_s}+m_{D_s}}\right)^2 H(w)\right],\\
H(w)&=&\frac{m_{H_s}+m_{D_s}}{m_{H_s}-m_{D_s}} \frac{h^-(w)}{h^+(w)}.
\end{eqnarray}
$h^\pm(1)$, $f^\pm(1)$ and $f_0(1)$ can then be extracted by doing a polynomial extrapolation in $w-1$.
\emph{In the following we will concentrate on estimating of ${\cal G}^{B_s \to D_s}(w=1)$}.

To obtain results at the $b$ quark mass (while trying to keep cut-off effects under control) we have decided to apply the strategy 
of step-scaling in masses \cite{Blossier:2009hg, Atoui:2013mqa} that was already adapted to the Wilson-Clover regularisation in \cite{Balasubramanian:2019net}. 
Steps in RGI heavy quark mass cannot be used at this stage of knowledge as far as the Wilon-Clover
regularisation is concerned. Indeed without a still unknown
$O(a^2)$ term in an improvement factor, the RGI quark mass becomes negative, hence unphysical, for quark masses
greater than $\sim 2 m_c$ and the lattice spacings at our disposal.

The idea is to consider ratios of ${\cal G}(1)$ at 2 consecutive heavy-strange meson masses $m_{H_s(i+1)}$ and $m_{H_s(i)}$ in the step-scaling mass chain:
\begin{equation}
\sigma^i_{\cal G}=\frac{{\cal G}(1, m_{H_s(i+1)})}{{\cal G}(1, m_{H_s(i)})},\quad  m_{H_s(i)}=\lambda^i m_{D_s}, \quad
\lambda=\left(\frac{m_{B_s}}{m_{D_s}}\right)^{1/K},
\end{equation}
where $K$ is the number of steps. Thus we have \begin{equation}{\cal G}(1, m_{B_s}) = {\cal G}(1, m_{D_s}) \times
\prod_{i=0}^{K-1} \sigma^i_{\cal G}, \end{equation} with
 \begin{eqnarray}
  {\cal G}(1, m_{D_s}) &=&\lim_{a \to 0\, , \, m_\pi \to m^{\rm phys}_\pi}  {\cal G}^{\rm lat}(1, m_{D_s}, a, m_\pi),\\
 \sigma^i_{\cal G} &=&\lim_{a \to 0\, , \, m_\pi \to m^{\rm phys}_\pi}  \Sigma^{i}_{\cal G}(a, m_\pi),
\end{eqnarray}
where $\Sigma_{\mathcal{G}}^i=\frac{{\cal G}^{\rm lat}(m_{H_s(i)})}{{\cal G}^{\rm lat}(m_{H_s(i-1)})}$
is the ratio of successive $\cal{G}$'s on the given ensemble.
By construction,  ${\cal G}(1, m_{D_s})$ is equal to 1. It will be a useful check that our numerical
data obey that constraint. Taking this into account, we have ${\cal G}(1, m_{B_s}) = \prod_{i=0}^{K-1} \sigma^i_{\cal G}$.

Concerning the decay $H_s \to D^*_s$, a convenient framework is again HQET.
Taking into account parities under $C$, $P$ and $T$ symmetries, the Lorentz structure decomposition of the hadronic matrix element $\langle D^*_s(p_{D^*_s},\epsilon^{(\lambda)})|A^\mu|H_s(p_{H_s})\rangle$ reads
\begin{equation}
\frac{\langle D^*_s(p_{D^*_s},\epsilon^{(\lambda)})|A^\mu|H_s(p_{H_s})\rangle}{2\sqrt{m_{H_s}m_{D^*_s}}}
 = \frac{i}{2}\epsilon^{*(\lambda)}_\nu[g^{\mu\nu}(1+w)h_{A_1} -v^\mu_{H_s} (v^\nu_{B_s} h_{A_2} +v^\nu_{D^ *_s}h_{A_3} )].
 \end{equation}
With
\begin{equation}
R^{\mu \rho}=\sum_\lambda \epsilon^{\rho(\lambda)}\langle D^*_s(p_{D^*_s},\epsilon^{(\lambda)})|A^\mu|H_s(p_{H_s})\rangle
\end{equation}
and the normalisation equation of the polarisation vectors of a vector meson of mass $m_V$ and momentum $k$,
\begin{equation}
\sum_\lambda \epsilon^{\mu (\lambda)}\epsilon^{*\nu (\lambda)} = -g^{\mu\nu} + \frac{k^\mu k^\nu}{m^2_V},
\end{equation}
we get
\begin{eqnarray}
\nonumber
\frac{-R^{\mu \rho}}{2\sqrt{m_{H_s}m_{D^*_s}}}&=&\frac{i}{2} \left[\left(g^{\mu \rho}-\frac{p^{\mu}_{D^*_s}p^\rho_{D^*_s}}{m^2_{D^*_s}}
\right)(1+w)h_{A_1}\right.\\
&&\left.\hspace{0,4cm}-\left(g^{\rho \nu}-\frac{p^{\rho}_{D^*_s}p^\nu_{D^*_s}}{m^2_{D^*_s}}\right)
\left(v_{\nu\, B_s}v^\mu_{H_s} h_{A_2}+v_{\nu\,B_s}v^\mu_{D^*_s}h_{A_3}\right)\right].
\end{eqnarray}
Taking the $H_s$ meson at rest and $i,j$ as spatial indices, we obtain
\begin{equation}
\frac{-R^{ij}}{2\sqrt{m_{H_s}m_{D^*_s}}}=\frac{i}{2} \left[\left(g^{ij}-\frac{p^i_{D^*_s}p^j_{D^*_s}}{m^2_{D^*_s}}
\right)(1+w)h_{A_1}+\frac{E_{D^*_s}p^i_{D^*_s}p^j_{D^*_s}}{m^3_{D^*_s}}h_{A_3}\right].
\end{equation}
It means that we can determine two form factors, $h_{A_1}$ and $h_{A_3}$.
With 
\begin{eqnarray}
\nonumber
A&=&\frac{1}{2\sqrt{m_{H_s}m_{D^*_s}}}\frac{i}{3} \sum_i R^{ii},\\ 
B&=&\frac{1}{2\sqrt{m_{H_s}m_{D^*_s}}}\frac{i}{6}\sum_{i \neq j} R^{ij}
\end{eqnarray}
and 
an isotropic momentum $\vec{p}_{D^*_s}=(\theta,\theta,\theta)$, we have
\begin{eqnarray}
\nonumber
A&=&\frac{1}{2}\left[\left(1+\frac{\theta^2}{m^2_{D^*_s}}\right)(1+w)h_{A_1}
-\frac{E_{D^*_s}\theta^2}{m^3_{D^*_s}} h_{A_3}\right],\\
B&=&\frac{1}{2}\left(\frac{\theta^2}{m^2_{D^*_s}}(1+w)h_{A_1}
-\frac{E_{D^*_s}\theta^2}{m^3_{D^*_s}} h_{A_3}\right).
\end{eqnarray}
Thus, we arrive at
\begin{eqnarray}
%\nonumber
h_{A_1}&=&\frac{2}{w+1}(A-B),\\
h_{A_3}&=&\frac{m^3_{D^*_s}}{E_{D^*_s}\theta^2}
\left[(w+1)\frac{\theta^2}{m^2_{D^*_s}} h_{A_1}-2B\right].
\end{eqnarray}
According to our step-scaling procedure, these quantities can finally be expressed as:

\begin{equation}
h_{A_1}(1, m_{B_s}) = h_{A_1}(1, m_{D_s}) \times
\prod_{i=0}^{K-1} \sigma^i_{h_{A_1}} \label{h_A1_product}
\end{equation}
\begin{eqnarray}
\nonumber
R_{X_3}(1,m_{B_s})&\equiv& \frac{h_{A_3}(1,m_{B_s})}{h_{A_1}(1, m_{B_s})}\\
& =& R_{X_3}(1, m_{D_s}) \times
\prod_{i=0}^{K-1} \sigma^i_{R_{X_3}},
\end{eqnarray}
with
 \begin{eqnarray}
 %\nonumber
  h_{A_1}(1, m_{D_s}) &=&\lim_{a \to 0\, ,\, m_\pi \to m^{\rm phys}_\pi}  h_{A_1}^{\rm lat}(1, m_{D_s}, a, m_\pi),\\
%\nonumber
 \sigma^i_Y &=&\lim_{a \to 0\, ,\, m_\pi \to m^{\rm phys}_\pi}  \Sigma^i_Y(a, m_\pi),\;\;Y=h_{A_1},R_{X_3}.
\end{eqnarray}

%%%%%%%%%%%%%%%%%%%%%%%%%%%%%%%%%%%%%%%
\section{\label{sec3}Computational setup}
%%%%%%%%%%%%%%%%%%%%%%%%%%%%%%%%%%%%%%%

\def\paperorletter{letter}

\begin{table}[t]
\begin{center}
\begin{tabular}{cccccccccc}
\hline
	\toprule
	id	&	$\beta$	&	$(\frac{L}{a})^3\times \frac{T}{a}$ 		&	$\kappa_{\rm sea}$		&	$a~(\rm fm)$	&	$\frac{m_{\pi}}{\rm MeV}$	& $Lm_{\pi}$ 	& $\#$cfgs&$\kappa_s$&$\kappa_c$
	\\[0.2cm]
\hline  
	\midrule
	A5	&	5.2		&	$32^3\times64$	& 	$0.13594$	& 	0.0751	  	& 	$333$	&4.1	& $198$&$0.135267$&$0.12531$\\  
	B6*	&			& 	$48^3\times96$	&	$0.13597$	& 			& 	$282$	&5.2	& $118$&$0.135257$&$0.12529$\\    
\hline 
	\midrule
	E5	&	5.3		&	$32^3\times64$	& 	$0.13625$	& 	0.0653	  	& 	$439$	&4.7	& $200$&$0.135777$&$0.12724$\\  
	F6	&			& 	$48^3\times96$	&	$0.13635$	& 			& 	$313$	&5	& $120$&$0.135741$&$0.12713$\\    
	F7	&			& 	$48^3\times96$	&	$0.13638$	& 			& 	$268$	&4.3	& $200$&$0.135730$&$0.12713$\\    
	G8*	&			& 	$64^3\times128$	&	$0.13642$	& 			& 	$194$	&4.1	& $100$&$0.135705$&$0.12710$\\    
\hline
	\midrule
	N6	&	$5.5$	&	$48^3\times96$	&	$0.13667$	& 	$0.0483$  	& 	$341$	&4	& $192$&$0.136250$&$0.13026$\\	
	O7	&		&	$64^3\times128$	&	$0.13671$	& 	 	& 	$269$	&4.2	& $60$&$0.136243$&$0.13022$ \\ 
	\bottomrule
\hline
\end{tabular} 
\end{center}
\caption{Parameters of the simulations: bare coupling $\beta = 6/g_0^2$, 
lattice resolution, hopping parameters of strange- and charm-quark masses, lattice spacing $a$ in physical units, pion mass, 
number of gauge configurations, bare strange and charm quark masses. (Ensembles with * were excluded from the extraction of $h_{A_1}$).}
\label{tabsim}
\end{table}

We have used in our analysis gauge field configuration ensembles generated by the CLS effort with $N_f=2$ non-perturbatively $\mathrm{O}(a)$ improved Wilson-Clover fermions 
\cite{Sheikholeslami:1985ij, Luscher:1996ug} and the plaquette gauge action \cite{Wilson:1974sk}. 
The parameter $\kappa_s$ was taken from \cite{Fritzsch_2012} and $\kappa_c$ from 
\cite{heitger2013charm}. They have been used in previous works eg. \cite{Blossier_2018,Balasubramanian:2019net}.

\begin{figure}[t]
\begin{center}
\begin{tabular}{ccc}
\includegraphics*[width=0.4\textwidth]{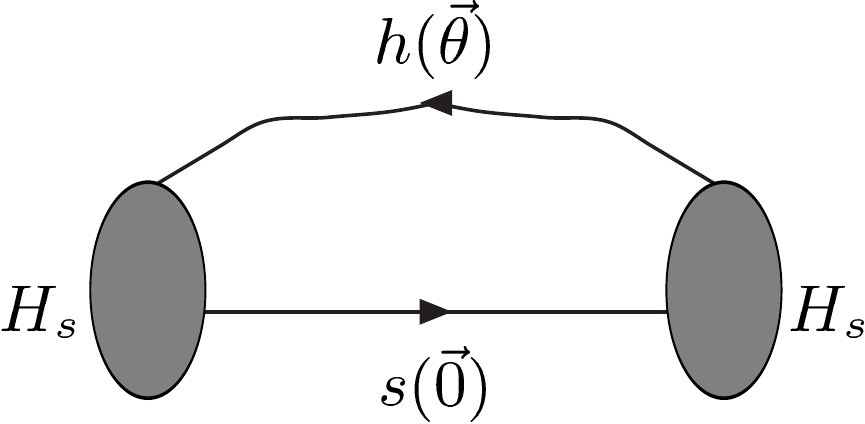}
&&
\includegraphics*[width=0.4\textwidth]{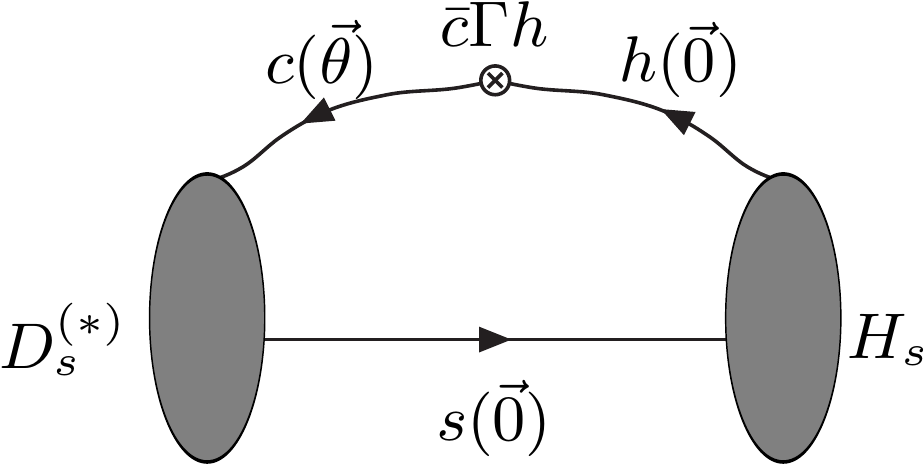}\\
\end{tabular}
\end{center}
\caption{\label{figkin}Kinematical configuration of the 2-pt. and 3-pt. correlation functions under study to extract
the $H_s \to D^{(*)}_s$ form factors: the strange quark is always set at rest, while the $h \to c$ current inherits its momentum from the charm quark boundary condition. $\Gamma$ specifies the Dirac structure in question.} 
\end{figure}

We have injected momenta to quarks by imposing
isotropic twisted-boundary conditions in space, i.e. $\psi(x+L\vec{e}_i)=e^{i \theta_i} \psi(x)$ with twisting angles $\theta_i$, $i=1,2,3$. The kinematics for 2-pt. and 3-pt. correlation function is depicted in Figure~\ref{figkin}, where strange, charm and heavy flavours are denoted by $s$, $c$ and $h$, respectively. 
Only the charm quark carries the non-zero momentum. To reduce some of the ${O}(a^2)$ effects we have averaged our results over positive and negative momenta. Six pairs $\pm \vec{\theta}$, in addition to the zero-momentum point, 
help us to extrapolate the various form factors associated with $H_s \to D_s$ and $H_s \to D^*_s$ from data available in the recoil region 
$1\leq w \lesssim 1.1$.

In Table~\ref{tabsim} we collect the parameter specifications of the ensembles. 
Three lattice spacings $a_{\beta=5.5}=0.04831(38)$ fm, $a_{\beta=5.3}=0.06531(60)$ fm, $a_{\beta=5.2}=0.07513(79)$ fm, 
where the scale setting was performed in   \cite{Lottini:2013rfa} via a fit in the chiral sector, 
are considered with pion masses in the range $[190 \,\rm MeV \,, 440\, \rm MeV]$. 
Heavy bare quark masses are the same as in \cite{Balasubramanian:2019net}.
 Statistical errors have been computed employing the ``$\Gamma$-method'' \cite{Wolff:2003sm}, 
 based on estimating autocorrelation functions\footnote{We are grateful to Fabian Joswig for having shared with us his Python implementation {\texttt{pyerrors}} of the $\Gamma$-method \url{https://github.com/fjosw/pyerrors}. }. As in \cite{Balasubramanian:2019net} all-to-all propagators were estimated stochastically with $U(1)$ spin-diluted walltime noise sources. We have also reduced the contamination by excited states on 2-pt. and 3-pt. correlators by solving a $4 \times 4$ Generalized Eigenvalue Problem (GEVP) with one local and 3 Gaussian smeared interpolating fields. In the mass step-scaling strategy outlined above, we set $K=6$.
In Table~\ref{tabtheta} we collect the twisting angles used to inject momenta to the charm quark. 
The phase shift was applied isotropically in all spatial directions.
\begin{table}[t]
\begin{center}
\begin{tabular}{c c} 
\hline
\toprule
id&$\theta \,[\pi / L]$\\
\hline \midrule
A5&\{0.000, $\pm$ 0.150, $\pm$ 0.300, $\pm$ 0.525, $\pm$ 0.750, $\pm$ 1.050, $\pm$ 1.275\}\\
\hline \midrule
B6&\{0.000, $\pm$ 0.163, $\pm$ 0.326, $\pm$ 0.571, $\pm$ 0.815, $\pm$ 1.141, $\pm$ 1.385\}\\ 
\hline \midrule
E5&\{0.000, $\pm$ 0.125, $\pm$ 0.250, $\pm$ 0.438, $\pm$ 0.625, $\pm$ 0.875, $\pm$ 1.062\}\\
\hline \midrule
F6&\{0.000, $\pm$ 0.188, $\pm$ 0.375, $\pm$ 0.656, $\pm$ 0.938, $\pm$ 1.312, $\pm$ 1.594\}\\
\hline \midrule
F7&\{0.000, $\pm$ 0.188, $\pm$ 0.375, $\pm$ 0.656, $\pm$ 0.938, $\pm$ 1.312, $\pm$ 1.594\}\\
\hline \midrule
G8&\{0.000, $\pm$ 0.250, $\pm$ 0.500, $\pm$ 0.875, $\pm$ 1.250, $\pm$ 1.750, $\pm$ 2.125\}\\
\hline \midrule
N6&\{0.000, $\pm$ 0.254, $\pm$ 0.507, $\pm$ 0.887, $\pm$ 1.262, $\pm$ 1.774, $\pm$ 2.155\}\\
\hline \midrule
O7&\{0.000, $\pm$ 0.338, $\pm$ 0.676, $\pm$ 1.183, $\pm$ 1.683, $\pm$ 2.365, $\pm$ 2.873\}\\
\hline \bottomrule
\end{tabular}
\end{center}
\caption{Twisted angles of the charm quark field boundary condition that we use to inject momenta at the $h \to c$ current.\label{tabtheta}}
\end{table}

Our analysis involves the following 2-pt. correlation functions evaluated on the gauge field ensembles at hand:
\begin{eqnarray}
\nonumber
C^{\theta, D_s}_{ij}(t) &=&\sum_{\vec{x}} \langle P^\theta_{cs,\, i}(t) P^{\dag\, \theta}_{cs,\, j}(0)\rangle,\\
\nonumber
C^{H_s}_{ij}(t) &=&\sum_{\vec{x}} \langle P_{hs,\, i}(t) P^{\dag}_{hs,\, j}(0)\rangle,\\
C^{\theta, D^*_s}_{ij}(t) &=&\frac{1}{3} \sum_{k=1}^3 \sum_{\vec{x}} \langle V^{\theta}_{cs,\, k,\, i}(t) V^{\dag\,\, \theta}_{cs,\, k,\, j}(0)\rangle,
\end{eqnarray}
where quark labels are as above, $i$ and $j$ specify smearing labels and spatial coordinates (summed over) are suppressed. The entering quark bilinears are defined as: $P_{hs}=\bar{\psi}_h\gamma^5 \psi_s$, $P^\theta_{cs}=\bar{\psi}^{\theta}_c\gamma^5 \psi_s$ and
$V^{\theta}_{cs,\,k}=\bar{\psi}^\theta_c \gamma_k \psi_s$.

Now we continue with the 3-pt. correlation functions:
\begin{eqnarray}
\nonumber
C^{\theta, PV^I_\mu P}_{ij}(t_i,t) &=&\sum_{\vec{x},\vec{y}} \langle P_{hs,\, i}(t_i) V^{I,\, \theta}_{hc,\, \mu}(t)P^{\dag\, \theta}_{cs,\,j}(0)\rangle,\\
C^{\theta, PA^I_kV_l}_{ij}(t_i,t) &=&\sum_{\vec{x},\vec{y}} \langle P_{hs,\, i}(t_i) A^{I,\,\theta}_{hc,\, \mu}(t)V^{\dag\, \theta}_{cs,l,\, j}(0)\rangle,
\end{eqnarray}
where the improved currents read $V^{I,\,\theta}_{hc,\, \mu}=\bar{\psi}_h\gamma_\mu \psi^\theta_c + a c_V \tilde{\partial}_\nu (\bar{\psi}_h\sigma_{\mu \nu} \psi^\theta_c)$, and $A^{I,\,\theta}_{hc}=\bar{\psi}^{\theta}_c\gamma_k \gamma^5 \psi_h + a c_A \tilde{\partial}_k (\bar{\psi}^{\theta}_c\gamma^5 \psi_h)$  with
$\tilde{\partial}_\mu f(x)\equiv \frac{1}{2} ({f(x+a\hat{\mu})-f(x-a\hat{\mu})})$. 

Accordingly, we have to solve the three GEVPs 
\begin{eqnarray}
\nonumber
C^{\theta, D_s}_{ij}(t) v^{\theta, D_s}_j(t,t_0)&=&\lambda^{\theta,D_s}(t,t_0)C^{\theta, D_s}_{ij}(t_0) v_j^{\theta, D_s} (t,t_0),\\
\nonumber
C^{H_s}_{ij}(t) v^{H_s}_j(t,t_0)&=&\lambda^{H_s}(t,t_0)C^{H_s}(t_0)_{ij} v^{H_s}_j(t,t_0),\\
C^{\theta, D^*_s}_{ij}(t) v^{\theta, D^*_s}_j(t,t_0)&=&\lambda^{\theta,D^*_s}(t,t_0)C^{\theta, D^*_s}_{ij}(t_0) v^{\theta, D^*_s}_j(t,t_0).
\end{eqnarray}
Then we project the 2-pt.\ and 3-pt.\ correlation functions onto the generalised eigenvector chosen at a given time $t_{\rm fix}$, $\vec{b}\equiv \vec{v}(t_{\rm fix},t_0)$:
\begin{eqnarray}
\nonumber
\tilde{C}^{\theta, D_s}(t)&\equiv&b^{\theta, D_s}_i C^{\theta, D_s}_{ij}(t) b^{\theta, D_s}_j,\\
\nonumber
\tilde{C}^{H_s}(t)&\equiv& b^{H_s}_i C^{H_s}_{ij}(t) b^{H_s}_j,\\
\nonumber
\tilde{C}^{\theta, D^*_s}(t)&\equiv&b^{\theta, D^*_s}_i C^{\theta, D^*_s}_{ij}(t) b^{\theta, D^*_s}_j,\\
\nonumber
\tilde{C}^{\theta, PV^I_\mu P}(t_i,t) &\equiv& b^{H_s}_i C^{\theta, PV^I_\mu P}_{ij}(t_i,t) b^{\theta, D_s}_j,\\
\label{eq:projcorr}
\tilde{C}^{\theta, PA^I_k V_l}(t_i,t) &\equiv&b^{H_s}_i C^{\theta, PA^I_k V_l}_{ij}(t_i,t) b^{\theta, D^*_s}_j.
\end{eqnarray}
The asymptotic behaviour of the 2-pt. functions is known to be given by %\textbf{??}:
\begin{eqnarray}
\nonumber
\tilde{C}^{\theta, D_s}(t) &\stackrel{t/a \gg 1}{\longrightarrow}& \frac{{\cal Z}^{\theta, D_s}}{2 E_{D_s}} \exp^{-E_{D_s} T/2} 
\cosh \left[E_{D_s}\left(t - \frac{T}{2}\right)\right],\\
\nonumber
\tilde{C}^{H_s}(t) &\stackrel{t/a \gg 1}{\longrightarrow}& \frac{{\cal Z}^{H_s}}{2 m_{H_s}} \exp^{-m_{H_s} T/2} 
\cosh \left[m_{H_s}\left(t - \frac{T}{2}\right)\right],\\
\tilde{C}^{\theta, D^*_s}(t) &\stackrel{t/a \gg 1}{\longrightarrow}& \frac{{\cal Z}^{\theta, D^*_s}}{2 E_{D^*_s}} \left(1+\frac{\theta^2}{m_{D^*_s}}\right)\exp^{-E_{D^*_s} T/2} 
\cosh \left[E_{D^*_s}\left(t - \frac{T}{2}\right)\right].
\label{eq:fitcorrel2pts}
\end{eqnarray}
 
Finally, the desired matrix elements may be computed from the large-time asymptotics of suitable ratios of the foregoing 2pt.- and 3pt.- correlation functions as 
\begin{eqnarray}
\nonumber
\frac{\tilde{C}^{\theta, PV^I_\mu P}(t_i,t)\sqrt{{\cal Z}^{\theta, D_s}{\cal Z}^{H_s}}}
{\tilde{C}^{\theta, D_s}(t)\tilde{C}^{H_s}(t_i-t)}&\stackrel{1 \ll t/a \ll (t_i-t)/a}{\longrightarrow}&\langle D_s(\theta) | V^I_\mu|H_s\rangle^{(b)},\\
\frac{\tilde{C}^{\theta, PA^I_k V_l}(t_i,t)\sqrt{{\cal Z}^{\theta, D_s}{\cal Z}^{H_s}}}
{\tilde{C}^{\theta, D^*_s}(t)\tilde{C}^{H_s}(t_i-t)}\left(1+\frac{\theta^2}{m^2_{D^*_s}}\right)&\stackrel{1 \ll t/a \ll (t_i-t)/a}{\longrightarrow}&R^{(b)}_{kl},
\label{eq:hadrmatelement}
\end{eqnarray}
where the label $(b)$ refers to bare hadronic matrix elements. Later, it will be convenient to note
\begin{equation}
\frac{1}{3} \sum_i \langle D_s|V^I_i|H_s\rangle^{(b)} \equiv 
\langle D_s|\vec{V}^I|H_s\rangle^{(b)},
\quad
\frac{1}{3} \sum_i R^{(b)}_{ii} \equiv \langle D^*_s|\vec{A}^I|H_s\rangle^{(b)}_{\parallel}.
\end{equation}
Upon mass dependent ${O}(a)$ improvement and multiplicative renormalisation the physical matrix elements are
\begin{eqnarray}
\nonumber
\langle D_s(\theta) | V^I_\mu|H_s\rangle&=&Z_V (1+b_V am^{VWI}_{hc}) \langle D_s(\theta) | V^I_\mu|H_s\rangle^{(b)},\\
R_{kl}&=&Z_A (1+b_A am^{VWI}_{hc}) R^{(b)}_{kl},
\end{eqnarray}
where the vector Ward identity mass $m^{VWI}_{hc}$ is related to the PCAC quark masses $m^{PCAC}_{hs}$, $m^{PCAC}_{cs}$
and $m^{PCAC}_{ss}$ by $m^{VWI}_{hc}=Z(m^{PCAC}_{hs}+m^{PCAC}_{cs} - m^{PCAC}_{ss})$. Perturbative and non-pertubative values for the coefficients  $Z_V$, $Z_A$, $c_V$, $c_A$, $b_V$,
$b_A$ and $Z$ are available from 
%\cite{DellaMorte:2008xb, DellaMorte:2005xgj,DellaMorte:2005aqe, 
\cite{DallaBrida:2018tpn, Sint:1997jx, Guagnelli:2000jw}.

%%%%%%%%%%%%%%%%%%%%%%%%%%%%%%%%%%%%%%%
\section{\label{sec4}Analysis}
%%%%%%%%%%%%%%%%%%%%%%%%%%%%%%%%%%%%%%%
%\flushleft

\def\paperorletter{letter}

\subsection{\label{subsec4a}Extraction of hadronic matrix elements}

\begin{figure}
    \center
    \includegraphics[width=0.48\textwidth]{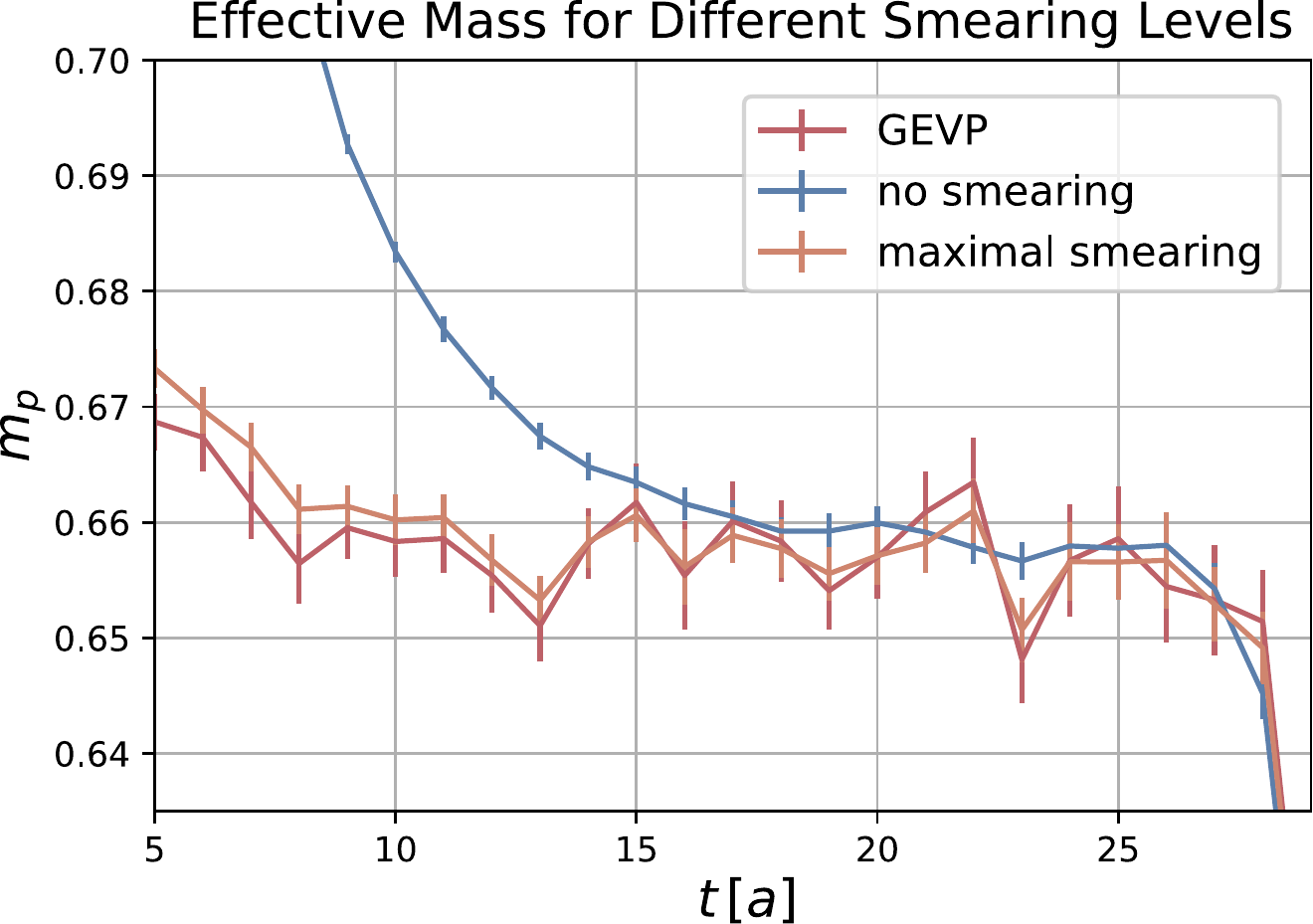} \hspace{0.05\textwidth}
    \includegraphics[width=0.40\textwidth]{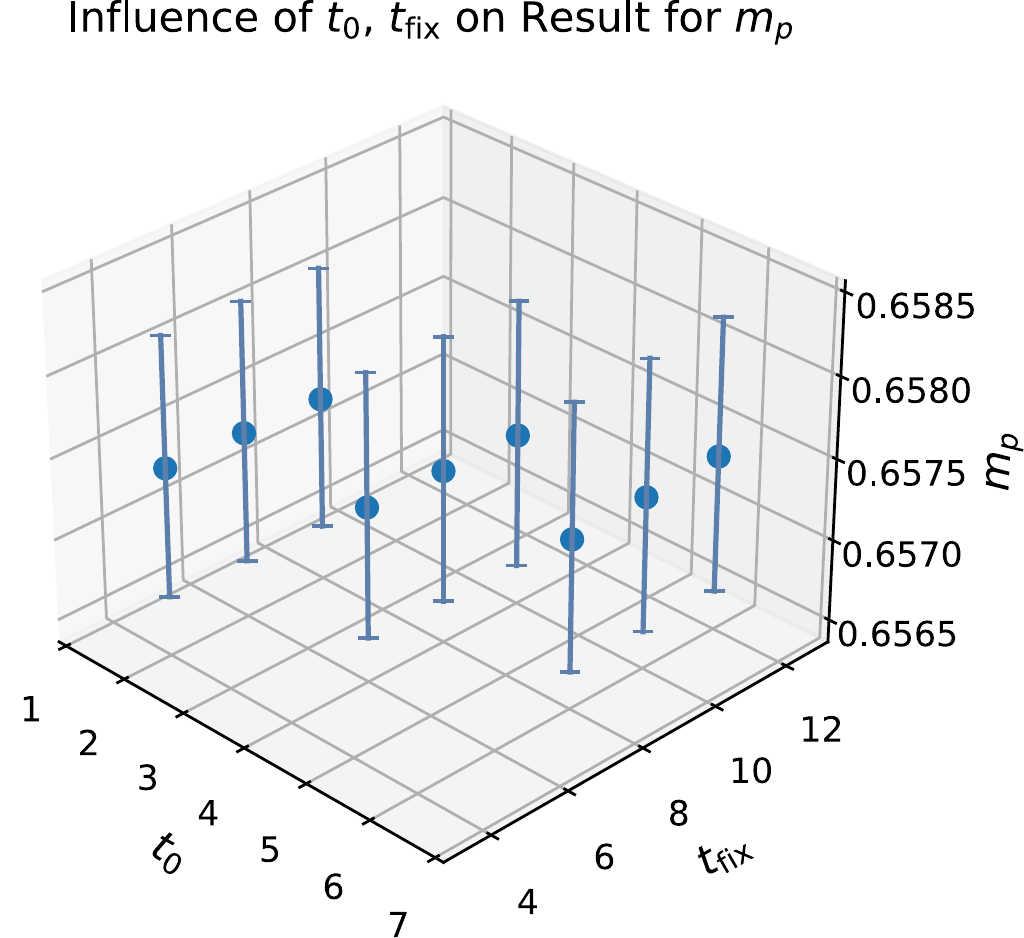}
    \caption{\label{GEVP-Figure}Visualisations for the generalised eigenvalue problem for the $D_s$ at
    rest on E5.
    \textit{Left}: Effective mass for the smearings [0,0],[80,80] and the GEVP projection.   
\textit{Right}: Result of the mass extraction with consistent plateau ranges for different GEVP parameters.}
\end{figure}

As described in the previous section, we have solved $4 \times 4$ GEVP systems, except for the most chiral ensemble G8 for which we had to restrict
the analysis to the $2 \times 2$ most smeared part of the correlators matrices, because the data are too noisy. We have set $t_0/a=3$ and $t_{\rm fix}/a=6$. 
We vary the parameters $t_{\rm fix}$, $t_0$ and the time ranges used to fit the 2-pt. correlation functions and ratios of 3-pt. and 2-pt. correlation functions, in order to 
estimate the systematic errors on the hadronic matrix elements.
In practice, we found that the alteration of  the parameters $t_0$ and $t_{\rm fix}$ does not influence the results significantly. 
Moreover we have observed that the GEVP solutions are in large part compatible with choosing the most smeared source for our interpolating fields.
%This, as well as the independence of $t_0$ and $t_{\rm fix}$ is shown in Figure \ref{GEVP-Figure} .
These findings are illustrated in Figure \ref{GEVP-Figure}.
%For the Ensemble G8 only the two sources with the largest smearing were available. Therefore the size of the 
%GEVP was reduced to $2\times 2$ on this Ensemble. 
The source-sink time separation $t_i$ is equal to $T/2$. 

We have obtained hadron masses $am_{H_{s}}$ by fitting the plateau 
of the effective mass data $am^{\rm eff}(t)$ coming from the generalised eigenvalues.\footnote{In practice, we got the eigenvalues, by projecting the smearing matrices with the eigenvectors.}
The fit range 
$[t_{\rm min}, t_{\rm max}]$ is chosen after a visual inspection of the plateau. The data are strongly correlated, therefore the error of the plateau is not much smaller than the error of the individual data points.
For this reason it was more important to exclude points outside the plateau than to maximise the available statistics. 
%{\bf XXX Does tmin is line
%with the criterion $\Delta m(tmin) < 3\delta m(tmin)$? XXX}
%$\Delta m(t)\equiv \exp (- \Delta E t_{\rm min})$, $\Delta E \sim$ 2 GeV.is the gap between the fifth  excited state heavy-strange hadron mass and the ground state hadron mass $M_{H_{i,s}}$. Its guesstimate is a way to assess the residual contamination by excited states that are not suppressed by the GEVP analysis. $t_{\rm max}$ is set by inspecting data statistical fluctuations. Those time ranges are then used to extract the ${\cal Z}$'s coefficients by fitting the projected 2-pt correlation functions $\tilde{C}^{\theta, D_s}$, $\tilde{C}^{H_s}$ and $\tilde{C}^{D^*_s}$.
The same is done to evaluate the hadronic matrix elements $\langle D_s|V^I_\mu|H_s\rangle^{(b)}$ and $R^{(b)}_{kl}$ along the formulae above.

%We show in Figure \ref{fig2} -- \ref{fig5} some effective masses of pseudoscalar and vector heavy-strange meson masses and hadronic matrix elements. Raw data are collected in the appendix.

\begin{figure}[t]
\begin{center}

\includegraphics*[width=1.0\textwidth]{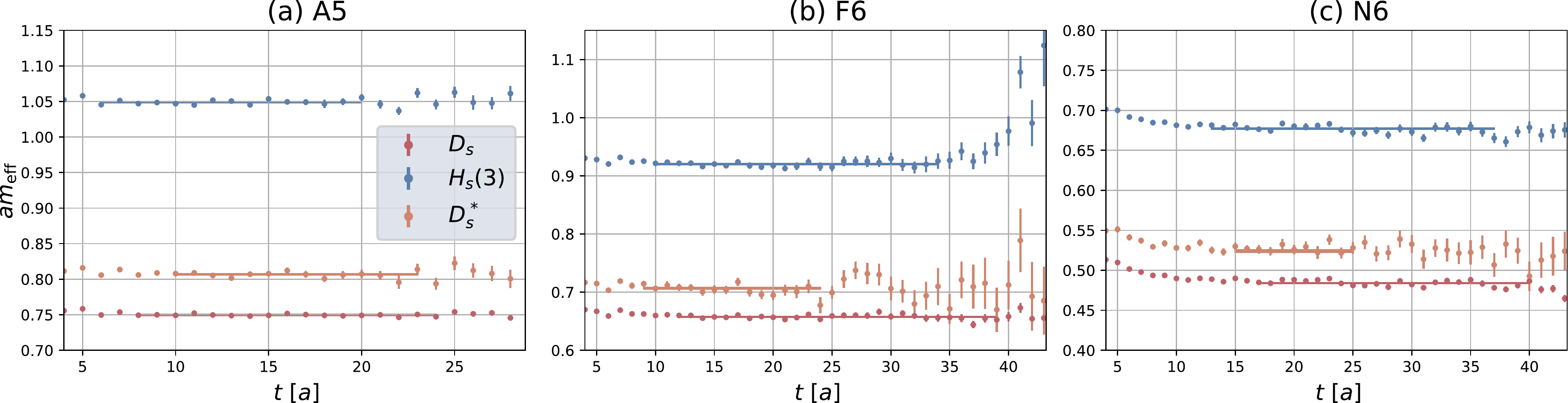}

\end{center}
\caption{\label{fig2} Effective masses obtained from the projected 2-pt.
correlators $\tilde{C}^{0,D_s}$, $\tilde{C}^{0,D^*_s}$ and $\tilde{C}^{H_s(3)}$,
together with the extracted mass. 
The data correspond to the ensemble A5 (a), F6 (b) and N6 (c).}

\end{figure}
\begin{figure}[]
\begin{center}

\includegraphics*[width=1.0\textwidth]{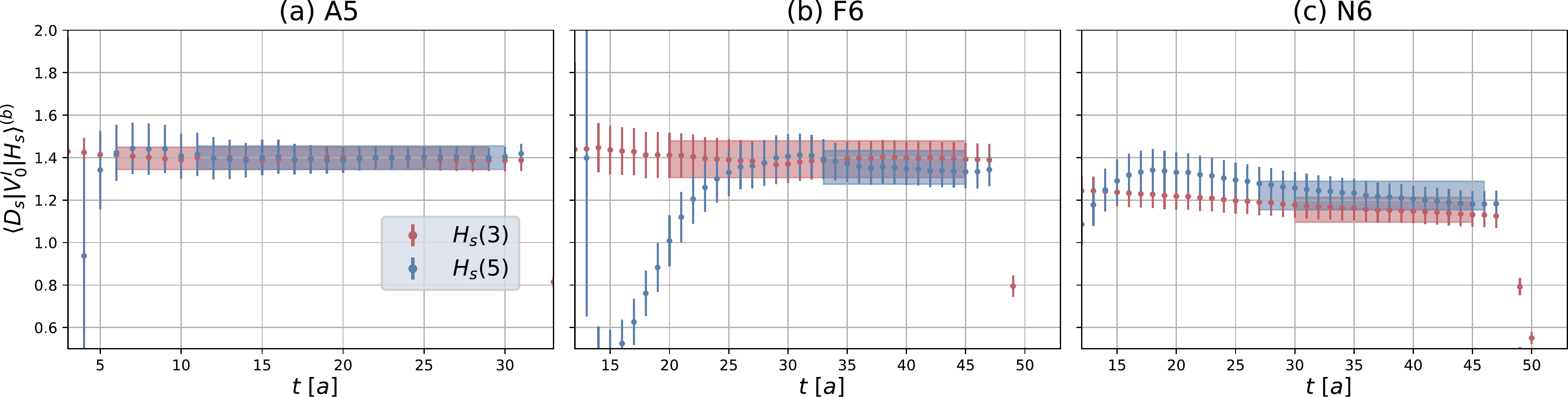}
\end{center}
\caption{\label{fig3} Bare hadronic matrix element $\langle D_s(\theta)|V^I_0|H_s(i=\{3,5 \})\rangle$
extracted from the fits according to eq.\,(\ref{eq:hadrmatelement}). 
The data correspond to the ensemble A5, $\theta=0.525 \pi/32$ (a), F6, $\theta=0.65625 \pi/48$ (b) and N6, $\theta=0.887 \pi/48$ (c).}
\end{figure}
\begin{figure}[t]
\begin{center}
\includegraphics*[width=1.0\textwidth]{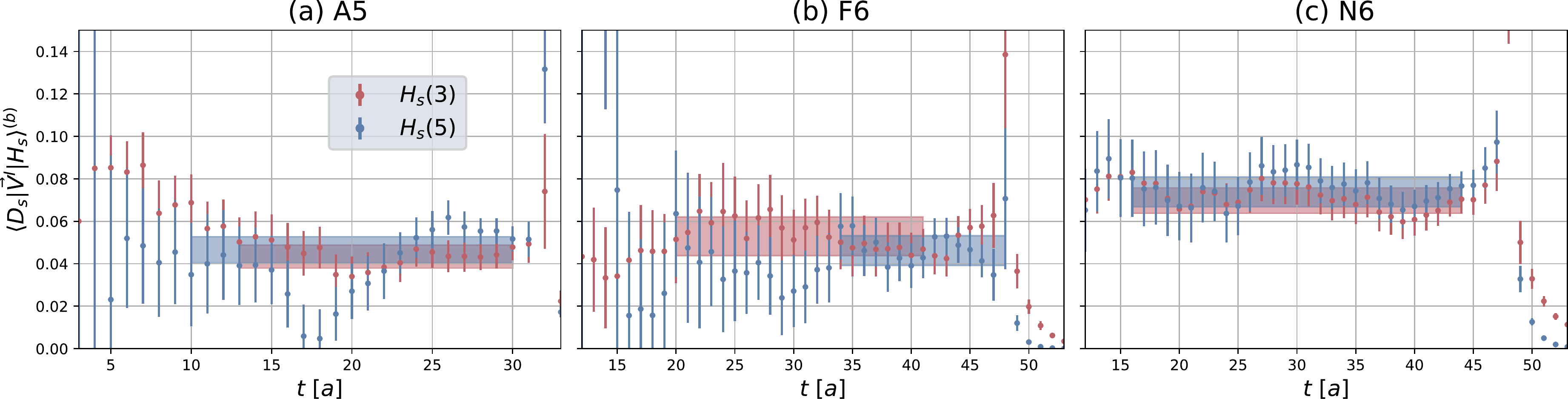}
\end{center}
\caption{\label{fig4} Bare hadronic matrix element $\langle D_s(\theta)|\vec{V}^I|H_s(i=\{3,5 \})\rangle^{(b)}$
extracted from the fits according to eq.\,(\ref{eq:hadrmatelement}). 
The data correspond to the ensemble A5, $\theta=0.525 \pi/32$ (a), F6, $\theta=0.65625 \pi/48$ (b) and N6, $\theta=0.887 \pi/48$ (c).}
\end{figure}

\begin{figure}[]
    \begin{center}
    \includegraphics*[width=1.0\textwidth]{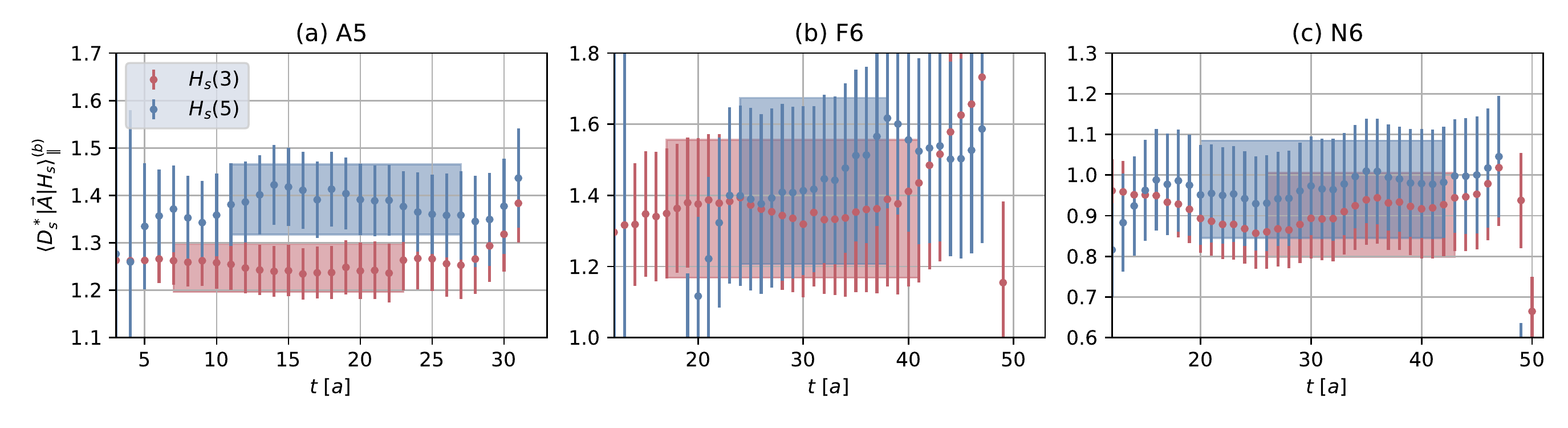}
    \end{center}
    \caption{\label{fig5} Bare hadronic matrix element 
    $\langle D^*_s(\theta)|\vec{A}|H_s(i=\{3,5 \})\rangle^{(b)}_{\parallel}$
   % $2 \sqrt{m_{H_s(i=3,5)} m_{D^*_s}}A^{(b)}(\theta, H_s(i=3,5)$
    extracted from the fits according to eq.\,(\ref{eq:hadrmatelement}). 
    The data correspond to the ensemble A5, $\theta=0.525 \pi/32$ (a), F6, $\theta=0.65625 \pi/48$ (b) and N6, $\theta=0.887 \pi/48$ (c).}
    \end{figure}
    In more detail, the extraction of the matrix elements involves the following elements
    \begin{center}
     
    \begin{itemize}
        \item[1] We apply symmetries, solve the GEVP and project the correlators following eq.\,(\ref{eq:projcorr}). 
        \item[2] The masses and energies, $am_{H_s}$ and $aE_{D_s^{(*)}}$, are extracted from a plateau of the effective mass of the projected correlator.
        \item[3]  The amplitudes ${\cal Z}^{\theta, H_s}$ and ${\cal Z}^{\theta, D^{(*)}_s}$ are obtained from a fit with eq.\,(\ref{eq:fitcorrel2pts}).
        \item[4] We divide the three-point-correlator by 
        $\frac{\sqrt{{\cal Z}^{\theta, H_s} {\cal Z}^{\theta, D^{(*)}_s}}}{2m_{H_s} E_{D_s^{(*)}}} e^{-E_{D_s^{(*)}}\frac{T}{2}} e^{-(m_{B_s}-E_{D_s^{(*)}})t}$
        and fit the resulting plateau to get $\langle D_s(\theta) | V^I_\mu|H_s\rangle^{(b)}$ or $R^{(b)}_{kk}$ 
        {(eq.\,(\ref{eq:hadrmatelement}) without the backward contributions in time.)}

    \end{itemize}
    \normalsize
\end{center}
Figures \ref{fig2} -- \ref{fig5} display some effective masses of pseudoscalar and vector heavy-strange mesons and hadronic matrix elements. Raw data are collected in the appendix.

    %\FloatBarrier

\subsection{\label{subsec4b}Extrapolation to the physical point}

\begin{figure}[t]
    \begin{center}
    \includegraphics*[width=0.32\textwidth]{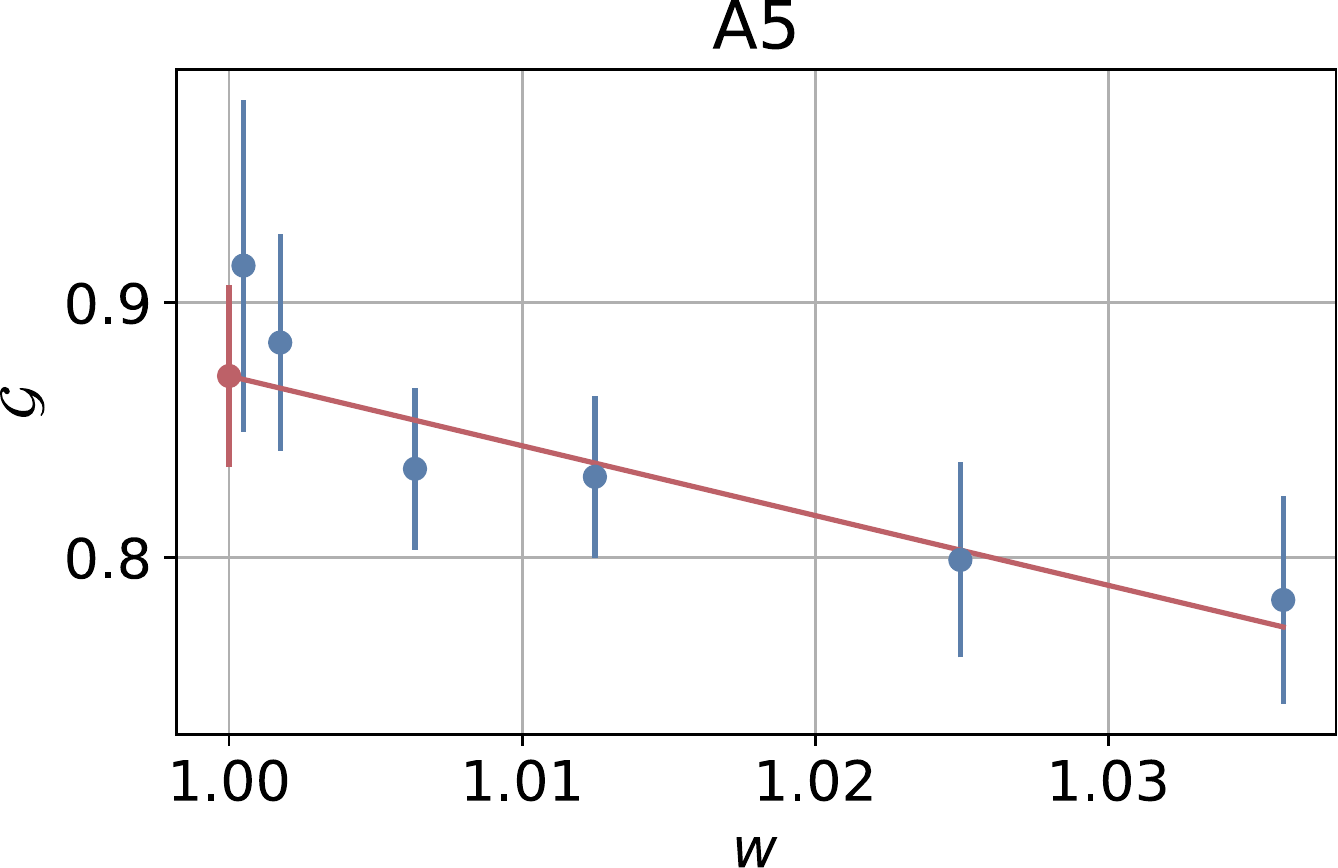}
    \includegraphics*[width=0.32\textwidth]{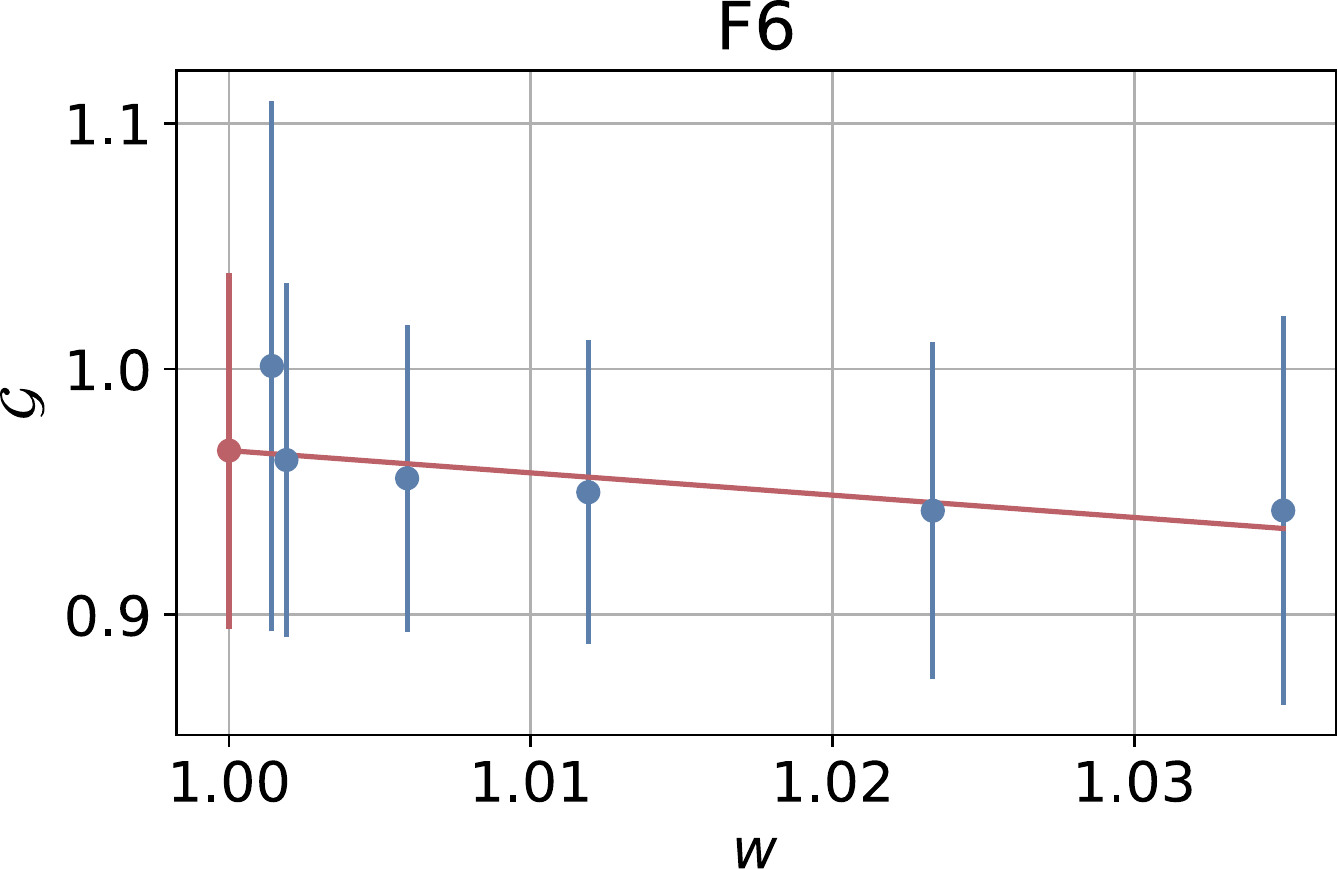}
    \includegraphics*[width=0.32\textwidth]{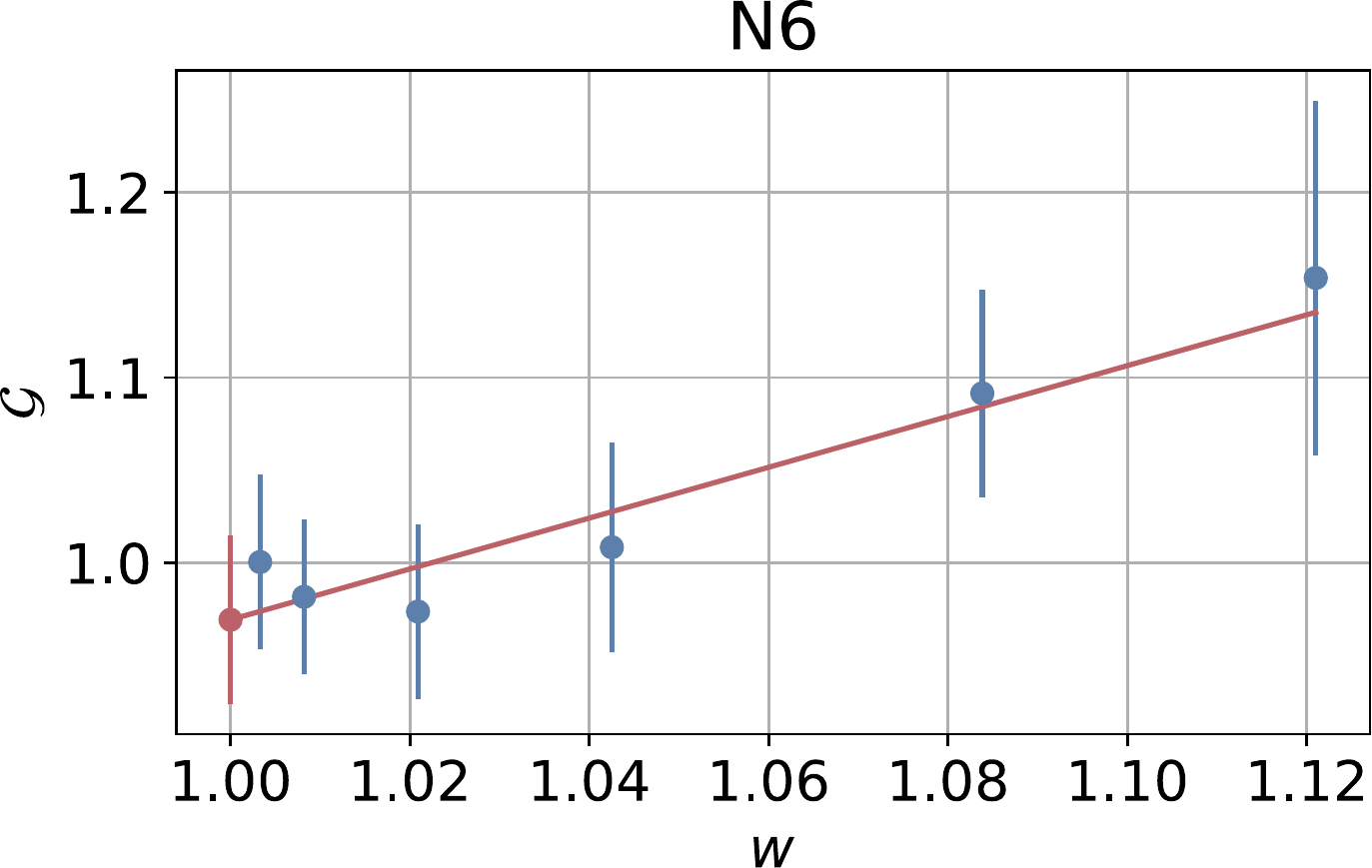}
    \end{center}
    \caption{\label{fig_wto1} Linear extrapolation of ${\cal{G}}(w)$ to the point of zero recoil ($w=1$).
    Examples given for three ensembles at $m_{H_s}=\lambda^2m_{D_s}$.}
    \end{figure}

Once we have obtained the hadronic matrix elements, we are in principle able to determine the heavy quark symmetry form factors $h^+$, $h^-$ and $h_{A_{1,3}}$. Unfortunately the statistical quality of our data is not sufficient to reliably calculate $h_{A_3}$.
Therefore we will restrict ourselves to the two former quantities at non zero recoil and on $h_{A_1}(w=1)$.
We have tried to extrapolate $h^+$ and $h^-$ to $w=1$ separately, and compute ${\cal G}(1)$ from it, as well as to extrapolate ${\cal G}(w)$ directly.
The two extrapolations are mostly compatible, because the range in $w$ is very small. We have performed linear and quadratic extrapolations in $(w-1)$. 
As quadratic fit coefficients are consistent with zero for almost all ensembles, we work with the result from the linear extrapolations for the remainder of the analysis.
In Figure \ref{fig_wto1} we show extrapolations of ${\cal G}$ to zero recoil for a selection of ensembles.
The next-to-last step is to extrapolate $\Sigma^i_{\cal G}=\frac{{\cal G}(1,a,m^2_\pi,m_{H_s(i)})}{{\cal G}(1,a,m^2_\pi,m_{H_s(i-1)})}$ to the physical point,
 where $i$ denotes the mass step-scaling step corresponding to $m_{H_s(i)}=\lambda^i m_{D_s}$.
We have used the following fit ansatz:
\begin{equation}
\Sigma^i_{\cal G}(w=1,a, m^2_\pi)=\Sigma^i_{{\cal G},0}
+ \Sigma^i_{{\cal G},1} \times (a/a_{\beta=5.3})^2
+\Sigma^i_{{\cal G},2} \times \left(\frac{m_\pi}{m^{\rm physical}_\pi}\right)^2.
\label{eqfitratioSigmaG}
\end{equation}
The fit parameters and $\chi^2/d.o.f.$ are collected in Table \ref{tabparafitextrapoSigmaG}.
\begin{table}[]
    \center
    \begin{tabular}{c c c c c}
        \hline \toprule  i& $\Sigma^i_{{\cal G},0}$ &  $\Sigma^i_{{\cal G},1}$ &  $\Sigma^i_{{\cal G},2}$ &$\chi^2/d.o.f.$\\
        \hline \midrule
        1 & 1.018(32) & -0.033(24) & 0.0006(32) & 0.46924\\ 
\hline \midrule
2 & 1.005(25) & -0.019(18) & 0.0000(27) & 0.35307\\
\hline \midrule
3 & 1.011(35) & 0.044(40) & -0.0057(43) & 0.56284\\
\hline \midrule
4 & 0.993(33) & 0.033(25) & -0.0028(29) & 0.91723\\ 
\hline \midrule
5 & 1.007(63) & -0.010(75) & 0.0009(75) & 1.09348\\ 
  \hline \bottomrule
\end{tabular}
\caption{\label{tabparafitextrapoSigmaG} Fit parameters for the extrapolation of $\Sigma^i_{\cal G}$. }
\end{table}

Parametrisations with additional terms were also studied.
 In particular, one might include a mistuning term $\left(1-\frac{m_{H_s}}{\lambda^i m_{Ds}}\right)$ to account for the fact that 
 the heavy meson masses are not tuned exactly equally on each ensemble. 

 However we decided to only include the terms in eq.\,(\ref{eqfitratioSigmaG})), because the data are so noisy that the fits can not reliably resolve these terms. In fact,  as seen in Table \ref{tabparafitextrapoSigmaG}, only very few non-constant 
 terms are statistically significant. Cut off effects are limited to 5\%, while a pion mass dependence is almost absent.
Then, we can write
\begin{equation}
\sigma^i_{\cal G}(m^{2,\, {\rm physical}}_\pi)=\Sigma^i_{\cal G}(w=1,a=0, m^2_\pi=m^{2,\, {\rm physical}}_\pi)=\Sigma^i_{{\cal G}_0}+\Sigma^i_{{\cal G}_2}.
\end{equation}

\begin{figure}
    \begin{center}
        \includegraphics[width=0.48\textwidth]{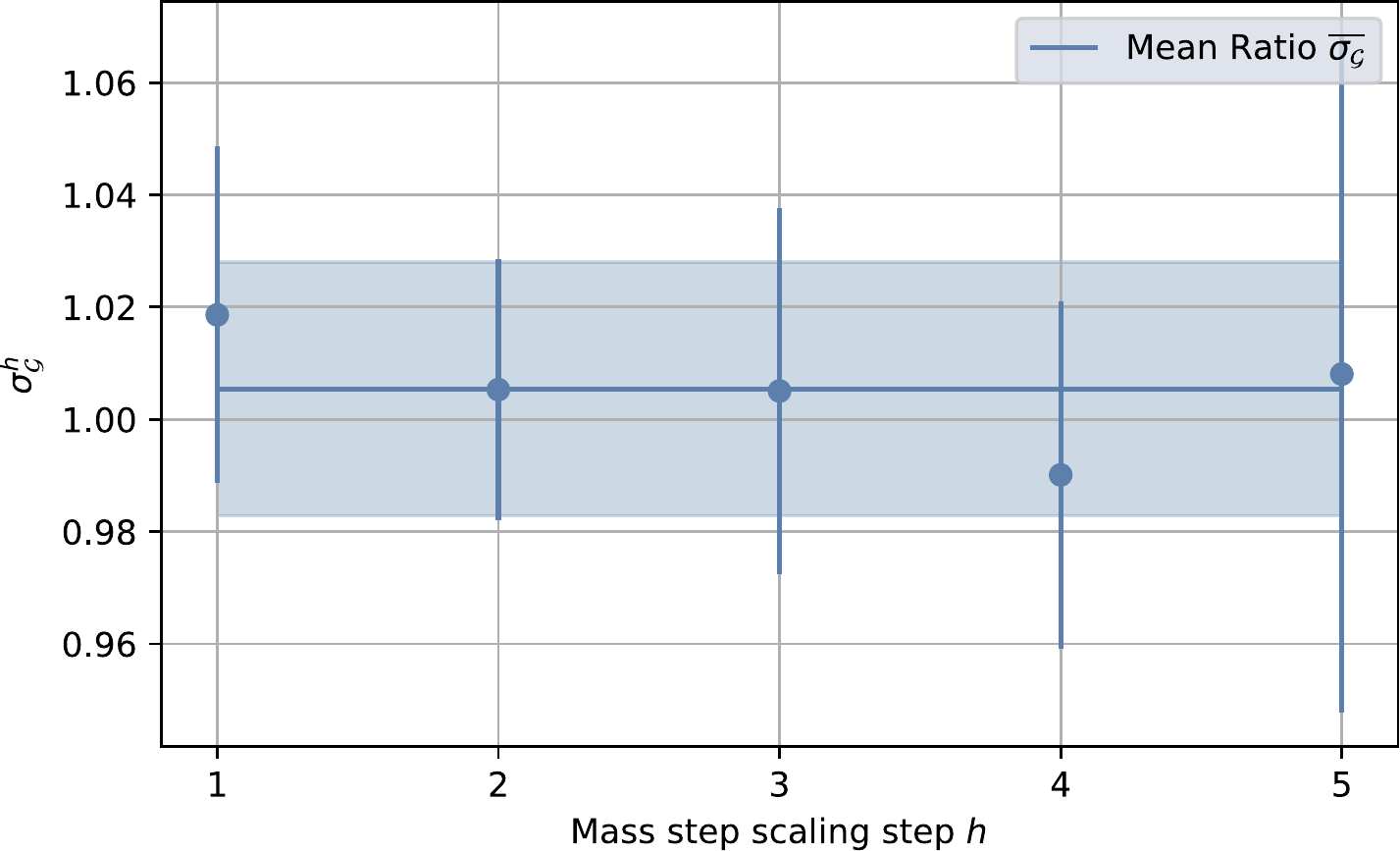}
        \includegraphics[width=0.48\textwidth]{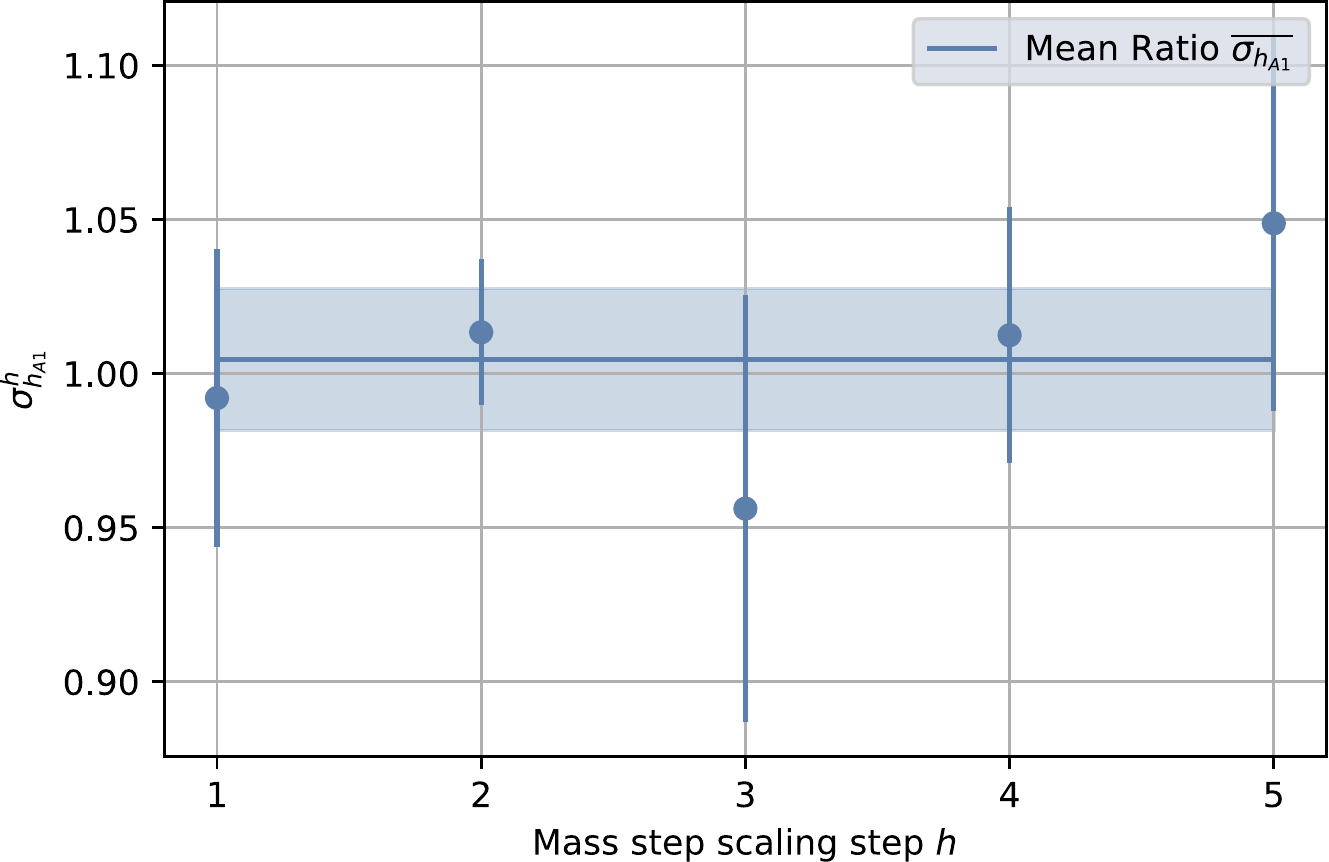}
        \caption{\label{fig_ratiofit} Average ratio of successive  $\cal G$ (\textit{left}) and $h_{A_1}$ (\textit{right}) at the physical point.}
    \end{center}
\end{figure}

Unfortunately, no clear trend in $m_{H_s}$ dependence is seen, as shown in the left panel of Figure \ref{fig_ratiofit}. 
That is why we have fitted the ratios $\sigma^i_{\cal G}$ by a constant $\overline{\sigma}_{\cal G}$. We get ${\overline{\sigma}_{\cal G}=1.005(23)}$. 

\begin{figure}
    \begin{center}
        \includegraphics[width=0.48\textwidth]{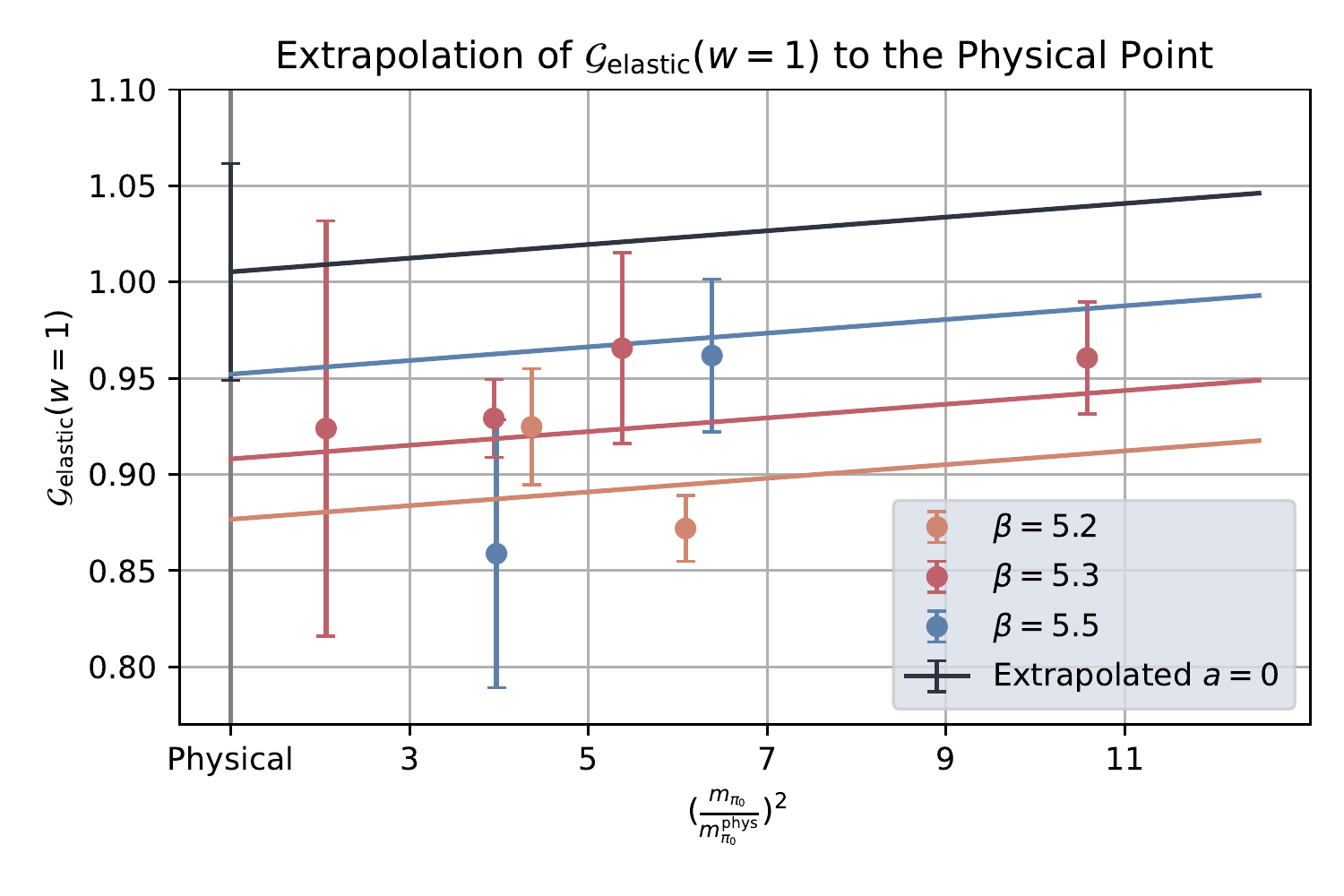}
        \includegraphics[width=0.48\textwidth]{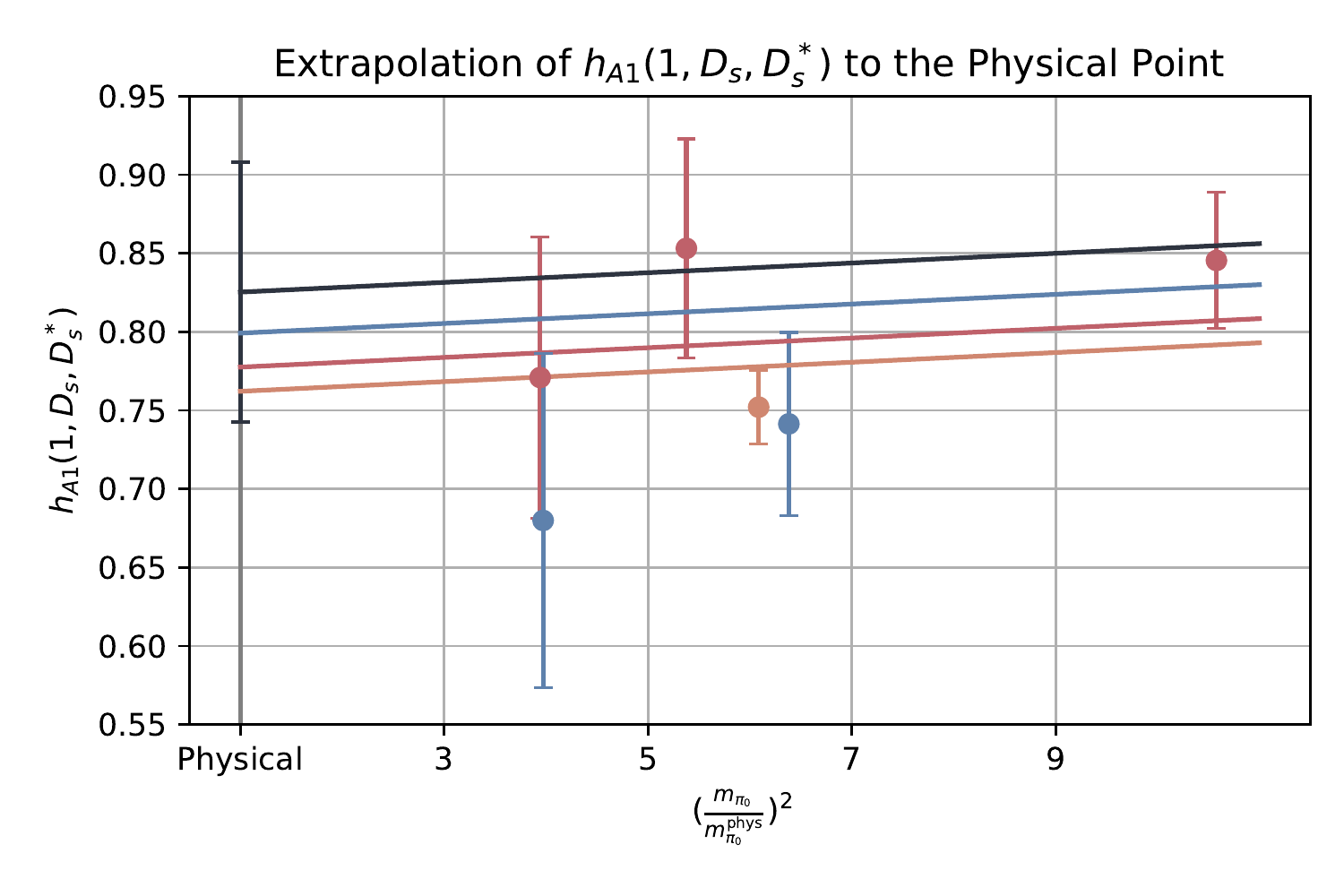}
        \caption{\label{fig_elastic} Visualisation of the combined fit eq.\,(\ref{elastic_fitfunc}) and extrapolation to the physical point for $\cal{G}_{\mathrm{elastic}}$ (\textit{left}) and $h_{A_1}(D_s,D_s^*)$ (\textit{right}). }
    \end{center}
\end{figure}

Moreover, we have checked, whether our data obey the constraint ${{\cal G}(1, m_{D_s}) =1}$. To do so, we have performed the
following extrapolation:
\begin{equation}
{\cal G}(1,a,m^2_\pi,m_{D_s})={\cal G}_0 + {\cal G}_1 \times (m^2_\pi / m^{2, {\rm physical}}_\pi)+{\cal G}_2 \times (a/a_{\beta=5.3})^2.
\label{elastic_fitfunc}
\end{equation}
The left panel of Figure \ref{fig_elastic} shows that our continuum extrapolation ${\cal G}(1, m_{D_s})$ is fully compatible with 1 within the limited statistics.
\begin{figure}
    \begin{center}
        \includegraphics[width=0.48\textwidth]{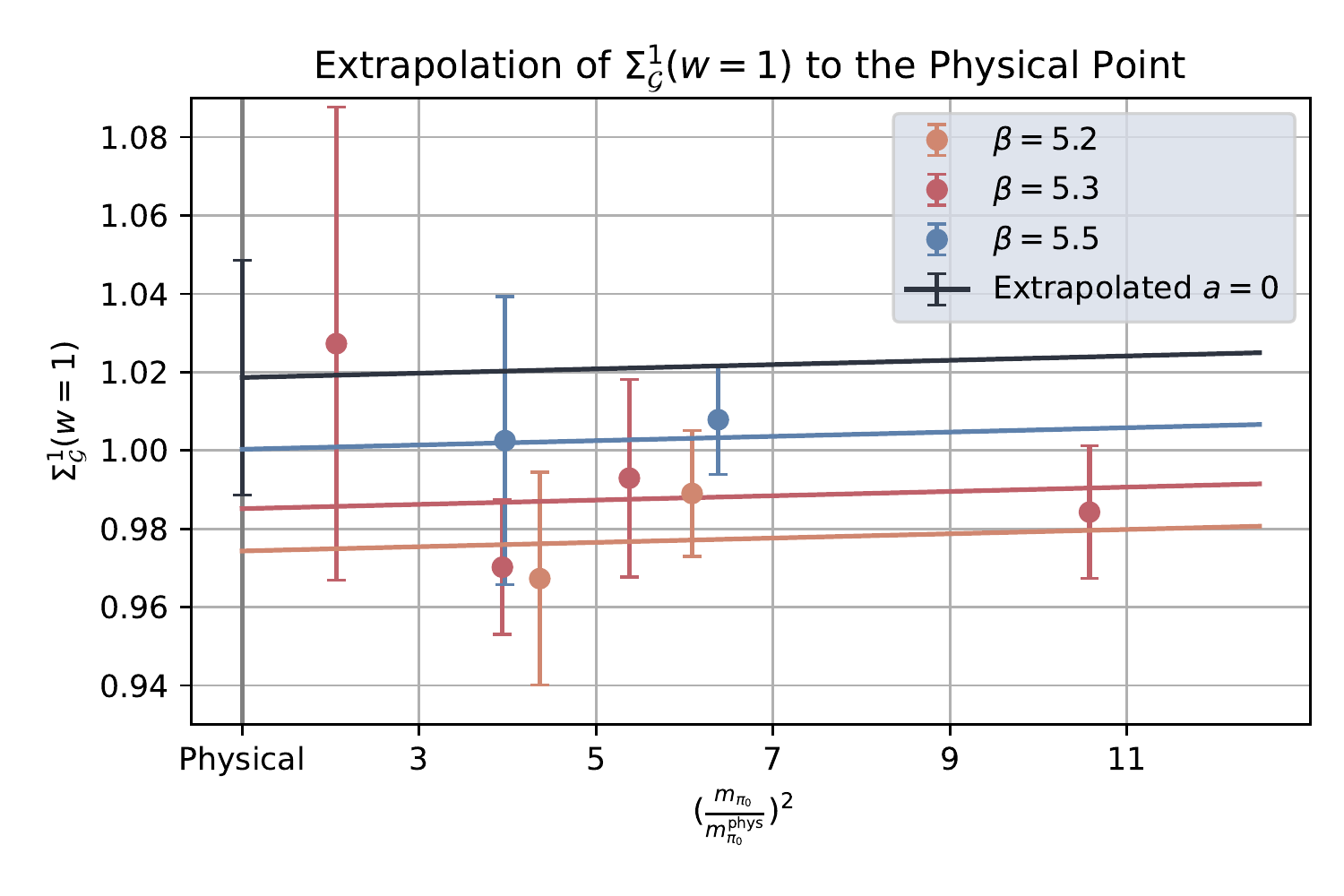}
        \includegraphics[width=0.48\textwidth]{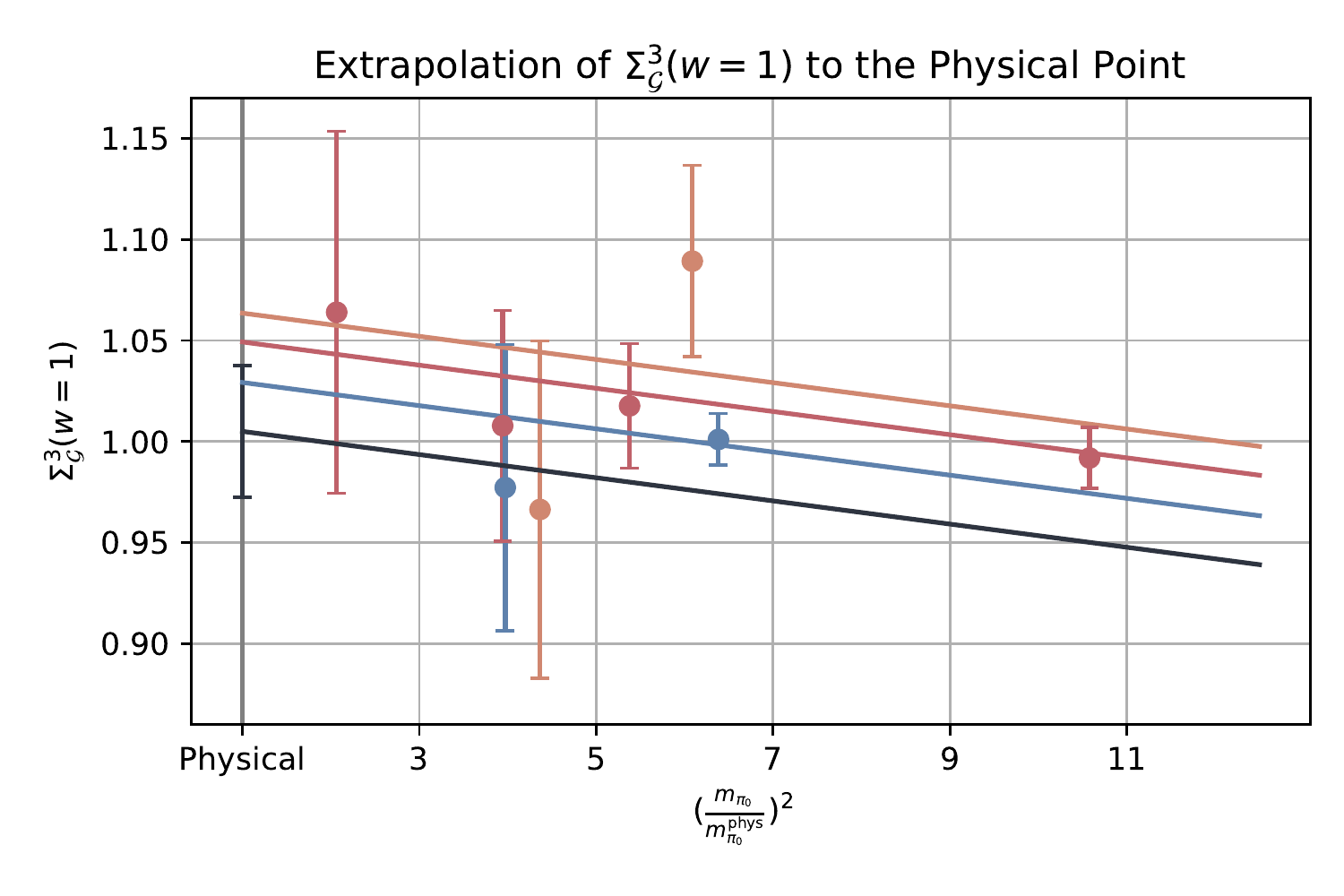}
        \caption{\label{fig_ratio_detail} Visualisation of the  extrapolation to the physical point of the ratio $\Sigma^i_{\cal{G}}(w=1)$ using eq.\,(\ref{eqfitratioSigmaG}) for two different mass step-scaling steps.}
    \end{center}
\end{figure}
As a final result for ${\cal G}$ we quote
\begin{equation}
    \label{final_result_G}
    {\cal G}^{B_s \to D_s}(w=1)= \overline{\sigma}_{\cal G}^{K}=\overline{\sigma}_{\cal G}^{6}=1.03(14).
\end{equation}
We have also performed a combined fit for the ratio using all data points simultaneously 
to explore the mass dependence and the effects of mistuning between the different 
ensembles:

    \begin{eqnarray}
\nonumber
\Sigma_{\cal G}(w=1,a, m_\pi,m_{H_s}) &=&   \Sigma_{{\cal G}_0}
    + \Sigma_{{\cal G}_1} \times (a/a_{\beta=5.3})^2
        +\Sigma_{{\cal G}_2} \times \left(\frac{m_\pi}{m^{ physical}_\pi}\right)^2 \\ 
	& +&\Sigma_{{\cal G}_3} \times (r_{\rm mis}-1)
    +\Sigma_{{\cal G}_4}\times \frac{m_{B_s}}{m_{H_s}},
    \label{eqfitratioSigmaGadvanced}
\end{eqnarray}
   where $r_{\rm mis}$ is defined as $\frac{am_{H_s}}{ \lambda^{i} a m_{Ds}}$ and $m_{H_s}$ denotes the physical meson mass. 
While the mistuning is not significant, $\Sigma_{{\cal G}_3}=0.69(71)$, an inclusion of this term makes the fit less stable. The last term does contribute 
$\Sigma_{{\cal G}_4}=-0.0248(91)$, but the ratios (analogous to Figure\,\ref{fig_ratiofit}) are still compatible with a constant. 
We obtain a final result for this method by computing 
\begin{equation}
    \prod\limits_{h=1}^6 \Sigma_{\cal G}(w=1,a=0, m_\pi=m_\pi^{\rm phys},m_{Ds} \lambda^{i})=
    1.03(14).
\end{equation}
Coincidently we get the same result as we get with the individual fits.

The extrapolation of $h_{A_1}$ to the physical point is completely analogous to that of ${\cal G}$.
We have decided to use individual fits as we did for the extrapolation of $G$.  
First we use the ansatz 
\begin{equation}
    {h_{A_1}}^{\rm lat}(w=1,a,m^2_\pi,m_{D_s})={h^0_{A_1}} + {h^1_{A_1}} (m_\pi / m^{ {\rm physical}}_\pi)^2+{h^2_{A_1}} (a/a_{\beta=5.3})^2
\end{equation}
to get ${h^{D_s \to D^*_s}_{A_1}}(1)$ as shown in the right panel of Figure \ref{fig_elastic}. Then the ratios of successive mass-step-scaling steps are fitted with

\begin{equation}
    \Sigma^i_{h_{A_1}}(1,a, m^2_\pi)=\Sigma^{i,0}_{{h_{A_1}}}
    \label{eqfitratioSigmahA1}+ \Sigma^{i,1}_{{h_{A_1}}} \times (a/a_{\beta=5.3})^2
    +\Sigma^{i,2}_{{h_{A_1}}} \times  \left(\frac{m_\pi}{m^{\rm physical}_\pi}\right)^2 .
    \end{equation}
The fit results are listed in Table \ref{tabparafitextrapoSigmahA1}.

\begin{table}[]
    \center
    \begin{tabular}{c c c c c}
        \hline \toprule  $i$& $\Sigma^{i,0}_{{h_{A_1}}}$ &  $\Sigma^{i,1}_{{h_{A_1}}}$ &  $\Sigma^{i,2}_{{h_{A_1}}}$ &$\chi^2/d.o.f.$\\
        \hline \midrule \\
        1 & 0.993(50) & -0.002(38) & -0.0011(24) & 0.56485\\ 
\hline \midrule
2 & 1.015(26) & -0.007(19) & -0.0018(26) & 0.47967\\ 
\hline \midrule
3 & 0.958(72) & 0.081(74) & -0.0022(63) & 1.08455\\ 
\hline \midrule
4 & 1.013(44) & 0.008(31) & -0.0005(39) & 0.11045\\ 
\hline \midrule
5 & 1.038(66) & -0.125(90) & 0.010(10) & 0.26381\\ 
 \\  \hline \bottomrule
\end{tabular}
\caption{ \label{tabparafitextrapoSigmahA1} Fit parameters for the extrapolation of $\Sigma^i_{h_{A_1}}$. }
\end{table}
Again, the pion mass dependence is very weak, below 1\%. It is less straightforward to conclude about cut off effects. They seem
to be more significant than for $\Sigma^i_{\cal G}$, as large as 12\%. 
As for $\Sigma^i_{\cal G}$, adding a mistuning term $\left(1-\frac{m_{H_s}}{\lambda^i m_{D_s}}\right)$ has made the fits unstable.
In the same way as before, from the ratios at the physical point and in the continuum limit, a fit to a constant $\overline{\sigma}_{h_{A_1}}$ of the ratios $\sigma^i_{h_{A_1}}$ is performed, because again no clear trend in $m_{H_s}$ is observed as shown in the right panel of Figure\,\ref{fig_ratiofit}. 
The result for $h_{A_1}$ can now be calculated according to eq.\,(\ref{h_A1_product})  
\begin{equation}
    h^{B_s \to D^*_s}_{A_1}(1)=h^{D_s \to D^*_s}_{A_1}(1) \times (\overline{\sigma}_{{h}_{A_1}})^K=0.825(83)\times(1.005(23))^{6}=0.85(16).
\end{equation}

\subsection{\label{subsec4c}Discussion}

To our knowledge, only three lattice results for ${\cal G}^{B_s \to D_s}(1)$ are quoted in the literature. The ETM Collaboration, in their analysis of ensembles with ${ N}_f=2$ twisted-mass fermions,  gets ${\cal G}^{B_s \to D_s}(1)=1.052(46)$, using the step-scaling
in mass method defined through RGI quark masses \cite{Atoui:2013mqa}. The HPQCD Collaboration has analysed ensembles with ${ N}_f=2+1$ staggered fermions, using the non-relativistic framework to regularise the $b$ quark. 
The result reads ${\cal G}^{B_s \to D_s}(1)=1.068(40)$ \cite{Monahan:2017uby}. 
They have also analysed ensembles with ${ N_f=2+1+1}$ staggered fermions, regularising the $b$ quark with the heavy HISQ action \cite{McLean:2019qcx}: they get  ${\cal G}^{B_s \to D_s}(1)=1.070(42)$\footnote{We thank Jonna Koponen for having provided us with the value, that is not explicitly quoted in \cite{McLean:2019qcx}.}.
Our result is in the same ballpark as the three previous lattice computations, with a significantly larger error.\footnote{The authors of \cite{Atoui:2013mqa} quote the same order of statistics for their matrix elements, but a much smaller error in their extrapolation.} A possible explanation is that we have set a large source-sink time separation $\grtsim\, 2$\,fm. This conservative choice helps to reduce the contamination from excited states with the caveat that the signal deteriorates faster for correlators build from heavy quarks.

Let us emphasize that a gain in statistics at each individual point of the step-scaling in mass strategy would have a significant impact to reduce the error on the final result, because the individual errors multiply with each other along the steps.

Three lattice results have been quoted in the literature for $h^{B_s \to D^*_s}_{A_1}(1)$ 
by HPQCD Collaboration at ${ N_f}=2+1$ and ${ N_f=2+1+1}$. 
Using non-relativistic $b$ quark, again, 
they get $h^{B_s \to D^*_s}_{A_1}(1)=0.883(40)$ \cite{Harrison:2017fmw} 
while, in the heavy HISQ framework, 
they get $h^{B_s \to D^*_s}_{A_1}(1)=0.9020(96)(90)$ \cite{McLean:2019sds} 
by a measurement direcly at zero recoil and $h^{B_s \to D^*_s}_{A_1}=0.875(45)$ \cite{Harrison:2021tol} 
by extrapolation to zero recoil.  
the authors of \cite{Harrison:2017fmw} find a tiny dependence on the spectator 
quark mass ($h^{B \to D^*}_{A_1}/h^{B_s \to D^*_s}_{A_1}=1.013(31)$). 
The FNAL/MILC Collaboration finds from another set of ensembles with ${ N_f=2+1}$ 
staggered fermions, $h^{B \to D^*}_{A_1}(1)=0.906(20)$, using the Fermilab action 
to regularize the $b$ quark \cite{Bailey:2014tva}. Two more recent preliminaries studies by this collaboration state $h^{B \to D^*}_{A_1}(1) \in [0.90, 0.95]$ \cite{Aviles-Casco:2017nge, Aviles-Casco:2019zop}. Thus there convincing evidence that $h^{B_s \to D^*_s}_{A_1}(1) \sim 0.9$. The same remark concerning our statistical error can be made for $h^{B_s \to D^*_s}_{A_1}$ as we did for ${\cal G}^{B_s \to D_s}(1)$ . The value of the ``elastic" point $h^{D_s \to D^*_s}_{A_1}(1)$ leads to the conclusion that heavy quark mass dependence on this form factor is smaller than 10\%, as it is for ${\cal G}^{H_s \to D_s}(1)$.

\begin{figure}
	\centering
	\includegraphics[width=0.95\textwidth]{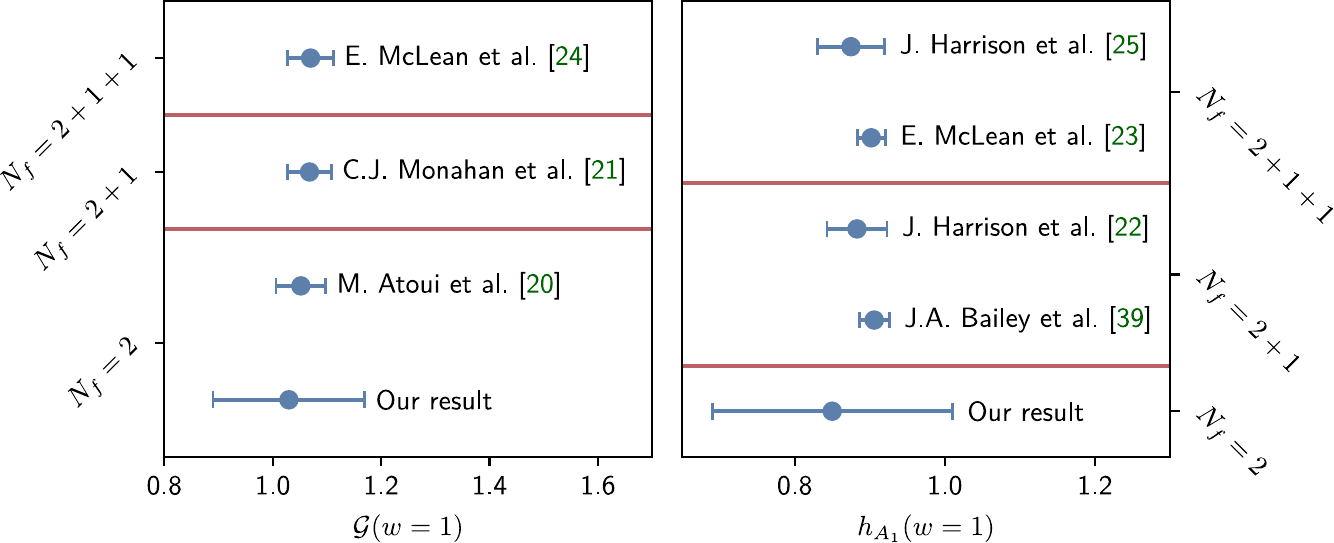}
	\caption{Comparison to previous results for $\mathcal{G}(w=1)$ (\textit{left}) and $h_{{\rm A}_1}$ (\textit{right}). \label{fig_comparison}}
\end{figure}

%%%%%%%%%%%%%%%%%%%%%%%%%%%%%%%%%%%%%%%
\section{\label{sec5}Conclusion}
%%%%%%%%%%%%%%%%%%%%%%%%%%%%%%%%%%%%%%%
%\flushleft

\def\paperorletter{letter}

In the present paper we have reported an ${ N_f=2}$ lattice QCD computation of the form factors ${\cal G}(1)$ and $h_{A_1}(1)$ 
associated to the $B_s \to D^{(*)}_s l \nu$ semileptonic decays.
For the purpose of extracting the underlying heavy-heavy hadronic matrix elements from suitable correlators, 
we have clearly demonstrated that solving the GEVP is beneficial to reduce their contamination
by excited states. Yet, only smearing $D^{(*)}_s$ and $B_s$ interpolating fields 
is already a good starting point. Step-scaling in masses helps to control cut-off effects originating from the heavy quark regularised on the lattice with the Wilson-Clover action. The twisted-boundary condition technique proves to be very helpful to extrapolate ${\cal G}(w)$ at zero recoil. 
It allows us to consider kinematics near-zero recoil so that a linear extrapolation in $(w-1)$ is fully
satisfactory. As we were conservative in setting the source-sink time separation larger than 2 fm in 3pt. correlation functions, we end up
with statistical errors $\sim 2.5\%$ in the heavy-strange meson masses sequence of form factor ratios. 
% This precision is not yet sufficient to be sensitive to the heavy quark mass dependence. 
% That is why in future works, especially when aiming at $B_s \to D^{(*)}_s$ and $B_c \to \eta_c (J/\psi)$ form factors using simulations at the physical point and at $N_f>2$, we are going to weaken our conservative choice for the source-sink time separation.
% Since the step-scaling in mass strategy tends to have a significant cost on lattice volumes as large as $100^3 \times 200$, % time will tell what the most convenient approach to reach a target error of $1\%$ on ${\cal G}(1)$ and $h_{A_1}(1)$ will be.
% reaching a target precision of 1\% via this approach might still be challenging.
This precision is not yet sufficient to be sensitive to the heavy quark
mass dependence and thus also entails the still considerably large overall
errors of the form factors from the present pilot study.

From the findings of our study, it can be deduced that a significant improvement
in precision can be expected from a reduced, less conservative choice for the
source-sink time separation and focusing on the most smeared sources, on top of
generally higher statistics of the individual ensembles of gauge field
configurations that enter the analysis.
However, rather than pursuing this for the two-flavour theory which is known to
have only limited impact on precision phenomenology, we intend to take these
aspects into account in a future calculation of $B_s\to D_s^{(*)}$
form factors using lattice QCD ensembles with $N_f=2+1$ dynamical quarks in the
medium term.
In fact, thanks again to large-scale simulations carried out by the CLS
(Coordinated Lattice Simulations) cooperation of researchers\footnote{%
https://www.zeuthen.desy.de/alpha/public-cls-nf21/},
a rich landscape of such $(2+1)$-flavour gauge field configuration ensembles
is already available \cite{Bruno:2014jqa,Bali:2016umi,Mohler:2017wnb}, which are
to be regarded as natural successors to the ones investigated in the present
work.
These ensembles employ non-perturbatively ${O}(a)$ improved Wilson-Clover
fermions, together with the tree-level Symanzik-improved gauge action, and cover
lattice spacings close to the continuum limit as well as the physical pion mass
point.
Since they also offer statistics substantially higher than underlying our
$N_f=2$ analysis, a computationally feasible application of the methods studied
(and its extensions suggested) here to the $N_f=2+1$ case, including additional
form factors of the $B_s$ system, now appears realistic.
Therefore, even though the cost of large-volume calculations via the step-scaling
in mass strategy remains challenging, an outcome of phenomenologically
interesting target precision of a few percent may be accomplished from this
approach on a reasonable time scale.

Finally, let us also mention the recently proposed, so-called stabilised Wilson
fermions as a new avenue for QCD calculations with (improved) Wilson-type
fermions \cite{Francis:2019muy}.
First simulations of lattice QCD with $2+1$ dynamical quark flavours within this
formulation~\cite{Cuteri:2022idv}, which amounts to replace the original clover
term with an exponentiated version of it, already hint at improved properties
such as, apart from stabilising simulations in large volumes towards small pion
masses, an overall good continuum scaling and milder relative quark mass
dependent cutoff effects.
Hence, this framework brings into sight another promising direction for
estimating strong interaction effects in precision weak interaction phenomenology
through lattice QCD that we are also envisaging for future applications of the
techniques explored in the present study.

\bibliography{paper}
\FloatBarrier
\section*{Acknowledgment}
This project is supported by Agence Nationale de la Recherche under the contract ANR-17-CE31-0019 (B.B., J.N. and S.LC.).
This work was granted access to the HPC resources of CINES and IDRIS (2018-A0030506808, 2019-A0050506808, 2020-A0070506808 and 2020-A0080502271) by GENCI.
We also gratefully acknowledge the computing time granted on SuperMUC-NG
(project ID {pn72gi}) by the Leibniz Supercomputing Centre of the
Bavarian Academy of Sciences and Humanities at Garching near Munich and
thank its staff for their support. The authors are grateful to the colleagues of the CLS effort for having provided the gauge field ensembles used in the present work.
This work is supported by the Deutsche Forschungsgemeinschaft (DFG) through the Research Training Group “GRK
2149: Strong and Weak Interactions – from Hadrons to Dark Matter” (J.N. and J.H.).

\begin{appendix}
%%%%%%%%%%%%%%%%%%%%%%%%%%%%%%%%%%%%%%%
%\newpage
\section*{\label{appendix}Appendix}

%%%%%%%%%%%%%%%%%%%%%%%%%%%%%%%%%%%%%%%

\def\paperorletter{letter}

Here we collect the raw data (masses, hadronic matrix elements and form factors) from our analysis.

%A5 !!!!!!!!!!!!!!!!!!!!!!!!!!!  A5 !!!!!!!!!!!!!!!!!!!!!!!!! A5 !!!!!!!!!!!!!!!!!!!

\begin{table}[h!] 
\center
\footnotesize
\begin{tabular}{rll}
\hline
   $\kappa_h$ & range   & $a m_p$       \\
\hline
     0.12531  & [8,24]  & 0.74953(64) \\
     0.121364 & [8,22]  & 0.88547(91) \\
     0.11604  & [6,20]  & 1.04906(68) \\
     0.109307 & [7,25]  & 1.23385(71) \\
     0.100407 & [8,20]  & 1.45746(88) \\
     0.089288 & [6,17]  & 1.71981(79) \\
\hline
\end{tabular}
\caption{Pseudoscalar masses on A5.}
\end{table}

\begin{table}[h!] 
\center
\footnotesize
\begin{tabular}{lrllllll}
\hline
 $\kappa_h$   &   $\theta$ & range    & $V_0^I$   & $V_0^{IR}$   & range    & $V_n^I$    & $V_n^{IR}$   \\
\hline
 0.12531      &      0     & [8, 29]  & 1.353(27) & 1.315(26)    & [17, 29] & 0.0032(15) & 0.0031(15)   \\
 0.12531      &      0.15  & [5, 28]  & 1.392(48) & 1.352(47)    & [15, 29] & 0.0185(55) & 0.0180(53)   \\
 0.12531      &      0.3   & [6, 28]  & 1.385(47) & 1.345(46)    & [10, 32] & 0.0263(45) & 0.0255(44)   \\
 0.12531      &      0.525 & [6, 27]  & 1.336(46) & 1.298(44)    & [10, 31] & 0.0443(47) & 0.0430(46)   \\
 0.12531      &      0.75  & [7, 27]  & 1.327(49) & 1.290(47)    & [10, 30] & 0.0631(53) & 0.0613(52)   \\
 0.12531      &      1.05  & [5, 26]  & 1.295(53) & 1.258(52)    & [18, 31] & 0.0859(73) & 0.0834(71)   \\
 0.12531      &      1.275 & [6, 28]  & 1.281(57) & 1.245(56)    & [12, 30] & 0.1043(78) & 0.1014(76)   \\
\hline
\end{tabular}
\caption{Improved and renormalised hadronic matrix elements $\langle D_s| V_\mu | H_s\rangle$ on A5.}
\end{table}

\begin{table} 
\center
\footnotesize
\begin{tabular}{lllll}
\hline
$\kappa_h$   & range    & $\langle D_s^*|\vec{A}^I|H_s\rangle^R_{\parallel}$      & $h_{A_1}$   & $\Sigma_{h_{A1}}$   \\
% $\kappa_h$   & range    & $2\sqrt{m_{H_s}m_{D^*_s}}A^R(\theta=0)$  & $h_{A_1}$   & $\Sigma_{h_{A_1}}$   \\
\hline
 0.12531      & [6, 24]  & 1.204(38) & 0.752(24)  & -                   \\
 \hline       &          &           &            &                     \\
 0.121364     & [7, 25]  & 1.229(44) & 0.739(26)  & 0.983(12)           \\
 \hline       &          &           &            &                     \\
 0.11604      & [7, 23]  & 1.247(51) & 0.732(30)  & 0.991(11)           \\
 \hline       &          &           &            &                     \\
 0.109307     & [8, 27]  & 1.400(64) & 0.820(37)  & 1.121(65)           \\
 \hline       &          &           &            &                     \\
 0.100407     & [11, 27] & 1.391(74) & 0.838(45)  & 1.022(14)           \\
 \hline       &          &           &            &                     \\
 0.089288     & [19, 31] & 1.206(81) & 0.777(52)  & 0.928(75)           \\
 %\hline       &          &           &            &                     \\
\hline
\end{tabular}
\caption{$h_{A_1}$ extracted on ensemble A5.}
\end{table}

\begin{table}[h!] 
\center 
\footnotesize 
\begin{tabular}{cll} 
\hline 
Scaling Step & $\mathcal{G}(w=1) $  & $\Sigma_{\mathcal{G}}$       \\ 
\hline 
$D_s$ & 0.872(17) & -\\
1 & 0.862(23) & 0.9890(10)\\
2 & 0.848(26) & 0.98289(81)\\
3 & 0.923(39) & 1.0893(29)\\
4 & 0.932(48) & 1.00921(92)\\
5 & 0.850(51) & 0.9124(46)\\
\hline 
\end{tabular}
\caption{$\mathcal{G}$ at zero recoil and $\Sigma_i=\frac{\mathcal{G}_i}{\mathcal{G}_{i-1}}$ on A5.} 
\end{table}

%B6 !!!!!!!!!!!!!!!!!!!!!!!!!!!  B6 !!!!!!!!!!!!!!!!!!!!!!!!! B6 !!!!!!!!!!!!!!!!!!!

\begin{table}[h!] 
\center
\footnotesize
\begin{tabular}{rll}
\hline
   $\kappa_h$ & range   & $a m_p$       \\
\hline
     0.12529  & [11,37] & 0.74884(34) \\
     0.121313 & [11,37] & 0.88526(48) \\
     0.116844 & [12,34] & 1.02281(57) \\
     0.109127 & [8,27]  & 1.23825(51) \\
     0.100172 & [7,23]  & 1.46300(53) \\
     0.088934 & [8,20]  & 1.72617(57) \\
\hline
\end{tabular}
\caption{Pseudoscalar masses on B6.}
\end{table}
 
\begin{table}[h!] 
\center
\footnotesize
\begin{tabular}{lrllllll}
\hline
 $\kappa_h$   &   $\theta$ & range    & $V_0^I$   & $V_0^{IR}$   & range    & $V_n^I$    & $V_n^{IR}$   \\
\hline
 0.12529      &    0       & [9, 40]  & 1.406(48) & 1.367(47)    & [14, 45] & 0.0045(19) & 0.0044(19)   \\
 0.12529      &    0.163   & [7, 44]  & 1.424(46) & 1.384(45)    & [12, 46] & 0.0091(37) & 0.0089(36)   \\
 0.12529      &    0.326   & [9, 44]  & 1.425(41) & 1.386(40)    & [9, 46]  & 0.0169(61) & 0.0164(60)   \\
 0.12529      &    0.5705  & [6, 42]  & 1.403(52) & 1.364(51)    & [14, 48] & 0.0270(70) & 0.0262(68)   \\
 0.12529      &    0.815   & [7, 40]  & 1.398(53) & 1.359(52)    & [12, 47] & 0.0377(68) & 0.0367(66)   \\
 0.12529      &    1.14095 & [15, 45] & 1.369(53) & 1.331(51)    & [9, 43]  & 0.0713(73) & 0.0693(71)   \\
 0.12529      &    1.38545 & [13, 45] & 1.361(53) & 1.323(51)    & [15, 42] & 0.0844(83) & 0.0821(81)   \\
\hline
\end{tabular}
\caption{Improved and renormalised hadronic matrix elements $\langle D_s| V_\mu | H_s\rangle$ on B6.}
\end{table}

\begin{table}[h!] 
\center 
\footnotesize 
\begin{tabular}{cll} 
\hline 
Scaling Step & $\mathcal{G}(w=1) $  & $\Sigma_{\mathcal{G}}$       \\ 
\hline 
$D_s$ & 0.925(30) & -\\
1 & 0.895(41) & 0.96727(89)\\
2 & 0.878(53) & 0.98133(78)\\
3 & 0.848(60) & 0.9664(31)\\
4 & 0.895(76) & 1.0551(10)\\
5 & 0.847(89) & 0.9459(45)\\
\hline 
\end{tabular}
\caption{$\mathcal{G}$ at zero recoil and $\Sigma_i=\frac{\mathcal{G}_i}{\mathcal{G}_{i-1}}$ on B6.} 
\end{table}

%E5 !!!!!!!!!!!!!!!!!!!!!!!!!!!  E5 !!!!!!!!!!!!!!!!!!!!!!!!! E5 !!!!!!!!!!!!!!!!!!!

\begin{table}[h!] 
\center
\footnotesize
\begin{tabular}{rll}
\hline
   $\kappa_h$ & range   & $a m_p$       \\
\hline
    0.12724   & [9,25]  & 0.65743(80) \\
    0.123874  & [8,27]  & 0.7765(10)  \\
    0.119457  & [9,27]  & 0.9166(11)  \\
    0.113638  & [7,27]  & 1.0835(12)  \\
    0.106031  & [7,22]  & 1.2813(13)  \\
    0.0965545 & [8,17]  & 1.5084(16)  \\
\hline
\end{tabular}
\caption{Pseudoscalar masses on E5.}
\end{table}

\begin{table}[h!] 
\center
\footnotesize
\begin{tabular}{lrllllll}
\hline
 $\kappa_h$   &   $\theta$ & range    & $V_0^I$   & $V_0^{IR}$   & range    & $V_n^I$    & $V_n^{IR}$   \\
\hline
 0.12724      &     0      & [5, 29]  & 1.325(43) & 1.261(41)    & [7, 32]  & 0.0037(20) & 0.0035(19)   \\
 0.12724      &     0.125  & [7, 27]  & 1.331(41) & 1.266(40)    & [5, 29]  & 0.0115(39) & 0.0110(37)   \\
 0.12724      &     0.25   & [5, 27]  & 1.340(42) & 1.275(40)    & [8, 29]  & 0.0232(41) & 0.0221(39)   \\
 0.12724      &     0.4375 & [6, 28]  & 1.365(43) & 1.299(41)    & [5, 30]  & 0.0427(44) & 0.0407(42)   \\
 0.12724      &     0.625  & [4, 29]  & 1.403(45) & 1.335(43)    & [9, 27]  & 0.0609(52) & 0.0580(49)   \\
 0.12724      &     0.875  & [4, 30]  & 1.465(51) & 1.394(48)    & [11, 30] & 0.0863(65) & 0.0821(62)   \\
 0.12724      &     1.0625 & [5, 28]  & 1.528(56) & 1.454(53)    & [12, 29] & 0.1078(75) & 0.1026(71)   \\
\hline
\end{tabular}
\caption{Improved and renormalised hadronic matrix elements $\langle D_s| V_\mu | H_s\rangle$ on E5.}
\end{table}

\begin{table} 
\center
\footnotesize
\begin{tabular}{lllll}
\hline
$\kappa_h$   & range    & $\langle D_s^*|\vec{A}^I|H_s\rangle^R_{\parallel}$      & $h_{A_1}$   & $\Sigma_{h_{A1}}$   \\
% $\kappa_h$   & range    & $2\sqrt{m_{H_s}m_{D^*_s}}A^R(\theta=0)$  & $h_{A_1}$   & $\Sigma_{h_{A_1}}$   \\
\hline
 0.12724      & [10, 23] & 1.212(63) & 0.845(43)  & -                   \\
 \hline       &          &           &            &                     \\
 0.123874     & [8, 20]  & 1.240(68) & 0.828(44)  & 0.9791(91)          \\
 \hline       &          &           &            &                     \\
 0.119457     & [8, 18]  & 1.271(78) & 0.821(50)  & 0.992(12)           \\
 \hline       &          &           &            &                     \\
 0.113638     & [9, 20]  & 1.310(91) & 0.833(57)  & 1.015(19)           \\
 \hline       &          &           &            &                     \\
 0.106031     & [10, 23] & 1.32(11)  & 0.847(67)  & 1.016(18)           \\
 \hline       &          &           &            &                     \\
 0.0965545    & [14, 29] & 1.30(11)  & 0.866(72)  & 1.023(22)           \\
 %\hline       &          &           &            &                     \\
\hline
\end{tabular}
\caption{$h_{A_1}$ extracted on Ensemble E5.}
\end{table}

\begin{table}[h!] 
\center 
\footnotesize 
\begin{tabular}{cll} 
\hline 
Scaling Step & $\mathcal{G}(w=1) $  & $\Sigma_{\mathcal{G}}$       \\ 
\hline 
$D_s$ & 0.961(29) & -\\
1 & 0.945(32) & 0.9842(11)\\
2 & 0.930(39) & 0.98338(98)\\
3 & 0.922(57) & 0.9919(10)\\
4 & 0.919(67) & 0.9971(11)\\
5 & 0.929(81) & 1.0100(13)\\
\hline 
\end{tabular}
\caption{$\mathcal{G}$ at zero recoil and $\Sigma_i=\frac{\mathcal{G}_i}{\mathcal{G}_{i-1}}$ on E5.} 
\end{table}

%F6 !!!!!!!!!!!!!!!!!!!!!!!!!!!  F6 !!!!!!!!!!!!!!!!!!!!!!!!! F6 !!!!!!!!!!!!!!!!!!!

\begin{table}[h!] 
\center
\footnotesize
\begin{tabular}{rll}
\hline
   $\kappa_h$ & range   & $a m_p$       \\
\hline
     0.12713  & [12,39] & 0.65756(70) \\
     0.1237   & [10,35] & 0.77899(69) \\
     0.119241 & [10,34] & 0.91989(79) \\
     0.113382 & [7,27]  & 1.08818(74) \\
     0.10593  & [8,22]  & 1.28429(94) \\
     0.096211 & [5,18]  & 1.51684(82) \\
\hline
\end{tabular}
\caption{Pseudoscalar masses on F6.}
\end{table}

\begin{table}[h!] 
\center
\footnotesize
\begin{tabular}{lrllllll}
\hline
 $\kappa_h$   &   $\theta$ & range    & $V_0^I$   & $V_0^{IR}$   & range    & $V_n^I$    & $V_n^{IR}$   \\
\hline
 0.12713      &    0       & [5, 43]  & 1.335(68) & 1.270(65)    & [13, 45] & 0.0057(33) & 0.0054(31)   \\
 0.12713      &    0.1875  & [6, 43]  & 1.344(68) & 1.279(65)    & [13, 45] & 0.0106(58) & 0.0100(55)   \\
 0.12713      &    0.375   & [9, 44]  & 1.332(71) & 1.268(68)    & [16, 44] & 0.0230(59) & 0.0219(57)   \\
 0.12713      &    0.65625 & [12, 44] & 1.337(72) & 1.272(69)    & [15, 44] & 0.0424(64) & 0.0403(61)   \\
 0.12713      &    0.9375  & [15, 44] & 1.343(76) & 1.278(73)    & [17, 41] & 0.0635(81) & 0.0604(77)   \\
 0.12713      &    1.3125  & [17, 46] & 1.358(88) & 1.293(84)    & [15, 44] & 0.0881(91) & 0.0838(87)   \\
 0.12713      &    1.59375 & [23, 46] & 1.37(11)  & 1.31(10)     & [14, 45] & 0.107(11)  & 0.102(10)    \\
\hline
\end{tabular}
\caption{Improved and renormalised hadronic matrix elements $\langle D_s| V_\mu | H_s\rangle$ on F6.}
\end{table}

\begin{table} 
\center
\footnotesize
\begin{tabular}{lllll}
\hline
$\kappa_h$   & range    & $\langle D_s^*|\vec{A}^I|H_s\rangle^R_{\parallel}$      & $h_{A_1}$   & $\Sigma_{h_{A1}}$   \\
% $\kappa_h$   & range    & $2\sqrt{m_{H_s}m_{D^*_s}}A^R(\theta=0)$  & $h_{A_1}$   & $\Sigma_{h_{A_1}}$   \\
\hline
 0.12713      & [9, 28]  & 1.22(10) & 0.853(70)  & -                   \\
 \hline       &          &          &            &                     \\
 0.1237       & [12, 28] & 1.31(14) & 0.872(94)  & 1.022(34)           \\
 \hline       &          &          &            &                     \\
 0.119241     & [17, 41] & 1.36(19) & 0.88(13)   & 1.010(56)           \\
 \hline       &          &          &            &                     \\
 0.113382     & [21, 42] & 1.40(23) & 0.89(15)   & 1.013(37)           \\
 \hline       &          &          &            &                     \\
 0.10593      & [24, 38] & 1.44(23) & 0.93(15)   & 1.04(24)            \\
 \hline       &          &          &            &                     \\
 0.096211     & [28, 40] & 1.37(27) & 0.92(18)   & 0.99(26)            \\
 %\hline       &          &          &            &                     \\
\hline
\end{tabular}
\caption{$h_{A_1}$ extracted on Ensemble F6.}
\end{table}

\begin{table}[h!] 
\center 
\footnotesize 
\begin{tabular}{cll} 
\hline 
Scaling Step & $\mathcal{G}(w=1) $  & $\Sigma_{\mathcal{G}}$       \\ 
\hline 
$D_s$ & 0.966(50) & -\\
1 & 0.959(58) & 0.9929(10)\\
2 & 0.967(72) & 1.0084(11)\\
3 & 0.984(86) & 1.0177(13)\\
4 & 0.911(82) & 0.9255(40)\\
5 & 1.104(98) & 1.2121(55)\\
\hline 
\end{tabular}
\caption{$\mathcal{G}$ at zero recoil and $\Sigma_i=\frac{\mathcal{G}_i}{\mathcal{G}_{i-1}}$ on F6.} 
\end{table}

%\clearpage

%F7 !!!!!!!!!!!!!!!!!!!!!!!!!!!  F7 !!!!!!!!!!!!!!!!!!!!!!!!! F7 !!!!!!!!!!!!!!!!!!!

\begin{table}[h!] 
\center
\footnotesize
\begin{tabular}{rll}
\hline
   $\kappa_h$ & range   & $a m_p$       \\
\hline
     0.12713  & [13,37] & 0.65511(28) \\
     0.123649 & [14,32] & 0.77744(47) \\
     0.119196 & [11,35] & 0.91865(36) \\
     0.11335  & [10,29] & 1.08532(41) \\
     0.105786 & [8,25]  & 1.28215(44) \\
     0.096689 & [7,23]  & 1.50077(43) \\
\hline
\end{tabular}
\caption{Pseudoscalar masses on F7.}
\end{table}

\begin{table}[h!] 
\center
\footnotesize
\begin{tabular}{lrllllll}
\hline
 $\kappa_h$   &   $\theta$ & range    & $V_0^I$   & $V_0^{IR}$   & range    & $V_n^I$    & $V_n^{IR}$   \\
\hline
 0.12713      &    0       & [8, 40]  & 1.272(26) & 1.212(25)    & [14, 45] & 0.0031(16) & 0.0030(15)   \\
 0.12713      &    0.1875  & [8, 44]  & 1.254(44) & 1.194(42)    & [6, 45]  & 0.0094(29) & 0.0089(27)   \\
 0.12713      &    0.375   & [9, 43]  & 1.253(44) & 1.194(42)    & [24, 39] & 0.0213(52) & 0.0203(49)   \\
 0.12713      &    0.65625 & [8, 43]  & 1.227(40) & 1.169(38)    & [23, 43] & 0.0307(57) & 0.0293(54)   \\
 0.12713      &    0.9375  & [7, 41]  & 1.226(41) & 1.168(39)    & [18, 45] & 0.0471(58) & 0.0448(55)   \\
 0.12713      &    1.3125  & [11, 43] & 1.340(69) & 1.276(65)    & [9, 43]  & 0.0804(51) & 0.0766(49)   \\
 0.12713      &    1.59375 & [13, 45] & 1.325(76) & 1.262(72)    & [12, 39] & 0.0971(65) & 0.0925(62)   \\
\hline
\end{tabular}
\caption{Improved and renormalised hadronic matrix elements $\langle D_s| V_\mu | H_s\rangle$ on F7.}
\end{table}

\begin{table} 
\center
\footnotesize
\begin{tabular}{lllll}
\hline
$\kappa_h$   & range    & $\langle D_s^*|\vec{A}^I|H_s\rangle^R_{\parallel}$      & $h_{A_1}$   & $\Sigma_{h_{A1}}$   \\
% $\kappa_h$   & range    & $2\sqrt{m_{H_s}m_{D^*_s}}A^R(\theta=0)$  & $h_{A_1}$   & $\Sigma_{h_{A_1}}$   \\
\hline
 0.12713      & [15, 25] & 1.175(41) & 0.825(29)  & -                   \\
 \hline       &          &           &            &                     \\
 0.123649     & [14, 30] & 1.211(52) & 0.813(35)  & 0.985(15)           \\
 \hline       &          &           &            &                     \\
 0.119196     & [14, 34] & 1.263(67) & 0.820(43)  & 1.009(15)           \\
 \hline       &          &           &            &                     \\
 0.11335      & [16, 34] & 1.226(83) & 0.782(53)  & 0.954(80)           \\
 \hline       &          &           &            &                     \\
 0.105786     & [19, 33] & 1.24(10)  & 0.794(65)  & 1.015(22)           \\
 \hline       &          &           &            &                     \\
 0.096689     & [22, 33] & 1.16(14)  & 0.771(90)  & 0.97(13)            \\
% \hline       &          &           &            &                     \\
\hline
\end{tabular}
\caption{$h_{A_1}$ extracted on Ensemble F7.}
\end{table}

\begin{table}[h!] 
\center 
\footnotesize 
\begin{tabular}{cll} 
\hline 
Scaling Step & $\mathcal{G}(w=1) $  & $\Sigma_{\mathcal{G}}$       \\ 
\hline 
$D_s$ & 0.929(20) & -\\
1 & 0.901(25) & 0.9702(11)\\
2 & 0.882(30) & 0.97820(88)\\
3 & 0.889(38) & 1.0078(40)\\
4 & 0.908(48) & 1.0214(10)\\
5 & 0.987(72) & 1.0875(58)\\
\hline 
\end{tabular}
\caption{$\mathcal{G}$ at zero recoil and $\Sigma_i=\frac{\mathcal{G}_i}{\mathcal{G}_{i-1}}$ on F7.} 
\end{table}

%G8 !!!!!!!!!!!!!!!!!!!!!!!!!!!  G8 !!!!!!!!!!!!!!!!!!!!!!!!! G8 !!!!!!!!!!!!!!!!!!!

\begin{table}[h!] 
\center
\footnotesize
\begin{tabular}{rll}
\hline
   $\kappa_h$ & range   & $a m_p$       \\
\hline
     0.1271   & [19,45] & 0.65660(86) \\
     0.123719 & [13,39] & 0.77676(98) \\
     0.11926  & [13,32] & 0.9179(11)  \\
     0.113447 & [12,33] & 1.0843(12)  \\
     0.105836 & [8,26]  & 1.28537(91) \\
     0.096143 & [5,24]  & 1.52492(90) \\
\hline
\end{tabular}
\caption{Pseudoscalar masses on G8.}
\end{table}

\begin{table}[h!] 
\center
\footnotesize
\begin{tabular}{lrllllll}
\hline
 $\kappa_h$   &   $\theta$ & range    & $V_0^I$   & $V_0^{IR}$   & range    & $V_n^I$    & $V_n^{IR}$   \\
\hline
 0.1271       &      0     & [11, 53] & 1.31(17)  & 1.25(16)     & [21, 53] & 0.0127(62) & 0.0121(59)   \\
 0.1271       &      0.25  & [9, 54]  & 1.28(15)  & 1.22(14)     & [9, 51]  & 0.0242(98) & 0.0230(93)   \\
 0.1271       &      0.5   & [8, 55]  & 1.28(15)  & 1.22(14)     & [14, 48] & 0.034(13)  & 0.032(13)    \\
 0.1271       &      0.875 & [11, 55] & 1.26(15)  & 1.20(14)     & [9, 48]  & 0.054(13)  & 0.051(12)    \\
 0.1271       &      1.25  & [11, 54] & 1.32(16)  & 1.26(15)     & [17, 49] & 0.070(16)  & 0.067(15)    \\
 0.1271       &      1.75  & [9, 52]  & 1.33(16)  & 1.27(15)     & [13, 48] & 0.090(21)  & 0.086(20)    \\
 0.1271       &      2.125 & [11, 55] & 1.35(18)  & 1.28(17)     & [12, 48] & 0.104(23)  & 0.099(22)    \\
\hline
\end{tabular}
\caption{Improved and renormalised hadronic matrix elements $\langle D_s| V_\mu | H_s\rangle$ on G8.}
\end{table}

\begin{table}[h!] 
\center 
\footnotesize 
\begin{tabular}{cll} 
\hline 
Scaling Step & $\mathcal{G}(w=1) $  & $\Sigma_{\mathcal{G}}$       \\ 
\hline 
$D_s$ & 0.92(11) & -\\
1 & 0.95(16) & 1.02726(97)\\
2 & 0.98(24) & 1.0322(13)\\
3 & 1.04(20) & 1.0640(15)\\
4 & 0.96(21) & 0.9256(17)\\
5 & 1.16(40) & 1.2036(32)\\
\hline 
\end{tabular}
\caption{$\mathcal{G}$ at zero recoil and $\Sigma_i=\frac{\mathcal{G}_i}{\mathcal{G}_{i-1}}$ on G8.} 
\end{table}

%N6 !!!!!!!!!!!!!!!!!!!!!!!!!!!  N6 !!!!!!!!!!!!!!!!!!!!!!!!! N6 !!!!!!!!!!!!!!!!!!!

\begin{table}[h!] 
\center
\footnotesize
\begin{tabular}{rll}
\hline
   $\kappa_h$ & range   & $a m_p$       \\
\hline
     0.13026  & [17,40] & 0.48394(56) \\
     0.127737 & [16,38] & 0.58010(66) \\
     0.124958 & [13,37] & 0.67722(66) \\
     0.121051 & [15,35] & 0.79993(87) \\
     0.115915 & [15,29] & 0.9483(10)  \\
     0.109399 & [10,23] & 1.12221(93) \\
\hline
\end{tabular}
\caption{Pseudoscalar masses on N6}
\end{table}

\begin{table}[h!] 
\center
\footnotesize
\begin{tabular}{lrllllll}
\hline
 $\kappa_h$   &   $\theta$ & range    & $V_0^I$    & $V_0^{IR}$   & range    & $V_n^I$    & $V_n^{IR}$   \\
\hline
 0.13026      &     0      & [5, 44]  & 1.019(38)  & 0.931(35)    & [13, 46] & 0.0051(24) & 0.0047(22)   \\
 0.13026      &     0.2535 & [17, 42] & 1.038(39)  & 0.947(36)    & [19, 45] & 0.0221(34) & 0.0202(31)   \\
 0.13026      &     0.507  & [23, 45] & 1.046(41)  & 0.955(38)    & [22, 46] & 0.0390(39) & 0.0356(36)   \\
 0.13026      &     0.887  & [16, 44] & 1.061(47)  & 0.969(43)    & [15, 43] & 0.0632(53) & 0.0577(49)   \\
 0.13026      &     1.2675 & [23, 45] & 1.110(59)  & 1.013(54)    & [16, 45] & 0.0895(67) & 0.0817(61)   \\
 0.13026      &     1.774  & [27, 48] & 1.221(71)  & 1.115(65)    & [14, 45] & 0.1343(78) & 0.1226(71)   \\
 0.13026      &     2.1545 & [31, 47] & 1.30(12)   & 1.19(11)     & [21, 44] & 0.161(15)  & 0.147(14)    \\
\hline
\end{tabular}
\caption{Improved and renormalised hadronic matrix elements $\langle D_s| V_\mu | H_s\rangle$ on N6.}
\end{table}

\begin{table} 
\center
\footnotesize
\begin{tabular}{lllll}
\hline
$\kappa_h$   & range    & $\langle D_s^*|\vec{A}^I|H_s\rangle^R_{\parallel}$      & $h_{A_1}$   & $\Sigma_{h_{A1}}$   \\
% $\kappa_h$   & range    & $2\sqrt{m_{H_s}m_{D^*_s}}A^R(\theta=0)$  & $h_{A_1}$   & $\Sigma_{h_{A_1}}$   \\
 \hline
 0.13026      & [14, 39] & 0.818(65) & 0.741(58)  & -                   \\
 \hline       &          &           &            &                     \\
 0.127737     & [27, 43] & 0.865(97) & 0.738(82)  & 0.996(46)           \\
 \hline       &          &           &            &                     \\
 0.124958     & [26, 43] & 0.90(10)  & 0.736(84)  & 0.9971(97)          \\
 \hline       &          &           &            &                     \\
 0.121051     & [15, 39] & 0.93(10)  & 0.732(81)  & 0.994(31)           \\
 \hline       &          &           &            &                     \\
 0.115915     & [20, 42] & 0.97(12)  & 0.739(91)  & 1.011(23)           \\
 \hline       &          &           &            &                     \\
 0.109399     & [25, 43] & 1.01(13)  & 0.765(97)  & 1.034(24)           \\
 %\hline       &          &           &            &                     \\
\hline
\end{tabular}
\caption{$h_{A_1}$ extracted on Ensemble N6.}
\end{table}

\begin{table}[h!] 
\center 
\footnotesize 
\begin{tabular}{cll} 
\hline 
Scaling Step & $\mathcal{G}(w=1) $  & $\Sigma_{\mathcal{G}}$       \\ 
\hline 
$D_s$ & 0.962(40) & -\\
1 & 0.969(44) & 1.00785(86)\\
2 & 0.967(46) & 0.99763(67)\\
3 & 0.968(54) & 1.00106(84)\\
4 & 0.968(64) & 0.9996(11)\\
5 & 0.967(62) & 0.9990(17)\\
\hline 
\end{tabular}
\caption{$\mathcal{G}$ at zero recoil and $\Sigma_i=\frac{\mathcal{G}_i}{\mathcal{G}_{i-1}}$ on N6.} 
\end{table}

%O7 !!!!!!!!!!!!!!!!!!!!!!!!!!!  O7 !!!!!!!!!!!!!!!!!!!!!!!!! O7 !!!!!!!!!!!!!!!!!!!

\begin{table}[h!] 
\center
\footnotesize
\begin{tabular}{rll}
\hline
   $\kappa_h$ & range   & $a m_p$       \\
\hline
     0.13022  & [15,44] & 0.48290(95) \\
     0.1279   & [17,42] & 0.5707(10)  \\
     0.124944 & [17,44] & 0.6729(12)  \\
     0.12091  & [16,41] & 0.7990(13)  \\
     0.11589  & [13,34] & 0.9433(12)  \\
     0.1094   & [8,28]  & 1.1156(13)  \\
\hline
\end{tabular}
\caption{Pseudoscalar masses on O7}
\end{table}

\begin{table}[h!] 
\center
\footnotesize
\begin{tabular}{lrllllll}
\hline
 $\kappa_h$   &   $\theta$ & range    & $V_0^I$   & $V_0^{IR}$   & range    & $V_n^I$     & $V_n^{IR}$   \\
\hline
 0.13022      &      0     & [15, 59] & 0.95(11)  & 0.871(98)    & [13, 62] & 0.0071(36)  & 0.0065(33)   \\
 0.13022      &      0.338 & [18, 60] & 0.944(80) & 0.863(73)    & [35, 63] & 0.0211(64)  & 0.0193(59)   \\
 0.13022      &      0.676 & [9, 59]  & 0.885(82) & 0.809(75)    & [36, 58] & 0.022(11)   & 0.0205(96)   \\
 0.13022      &      1.183 & [42, 61] & 0.848(92) & 0.776(84)    & [26, 62] & 0.049(11)   & 0.0446(97)   \\
 0.13022      &      1.683 & [43, 61] & 0.79(14)  & 0.72(13)     & [34, 64] & 0.061(14)   & 0.056(13)    \\
 0.13022      &      2.365 & [34, 62] & 0.79(20)  & 0.72(18)     & [49, 64] & 0.096(37)   & 0.088(34)    \\
 0.13022      &      2.873 & [28, 63] & 0.82(26)  & 0.75(24)     & [38, 62] & 0.061(38)   & 0.056(34)    \\
\hline
\end{tabular}
\caption{Improved and renormalised hadronic matrix elements $\langle D_s| V_\mu | H_s\rangle$ on O7.}
\end{table}

\begin{table} 
\center
\footnotesize
\begin{tabular}{lllll}
\hline
$\kappa_h$   & range    & $\langle D_s^*|\vec{A}^I|H_s\rangle^R_{\parallel}$      & $h_{A_1}$   & $\Sigma_{h_{A1}}$   \\
% $\kappa_h$   & range    & $2\sqrt{m_{H_s}m_{D^*_s}}A^R(\theta=0)$  & $h_{A_1}$   & $\Sigma_{h_{A_1}}$   \\
\hline
 0.13022      & [8, 31]  & 0.74(12) & 0.68(11)   & -                   \\
 \hline       &          &          &            &                     \\
 0.1279       & [7, 30]  & 0.75(15) & 0.65(13)   & 0.949(57)           \\
 \hline       &          &          &            &                     \\
 0.124944     & [33, 47] & 0.90(24) & 0.74(20)   & 1.15(16)            \\
 \hline       &          &          &            &                     \\
 0.12091      & [28, 49] & 1.08(28) & 0.86(22)   & 1.15(16)            \\
 \hline       &          &          &            &                     \\
 0.11589      & [28, 48] & 1.15(32) & 0.89(25)   & 1.039(48)           \\
 \hline       &          &          &            &                     \\
 0.1094       & [33, 46] & 2.06(66) & 1.58(51)   & 1.77(88)            \\
 \hline       &          &          &            &                     \\
\hline
\end{tabular}
\caption{$h_{A_1}$ extracted on Ensemble O7}
\end{table}

\begin{table}[h!] 
\center 
\footnotesize 
\begin{tabular}{cll} 
\hline 
Scaling Step & $\mathcal{G}(w=1) $  & $\Sigma_{\mathcal{G}}$       \\ 
\hline 
$D_s$ & 0.859(70) & -\\
1 & 0.861(76) & 1.00251(33)\\
2 & 0.846(84) & 0.98217(34)\\
3 & 0.826(91) & 0.97726(68)\\
4 & 0.80(10) & 0.96533(41)\\
5 & 0.79(13) & 0.98804(97)\\
\hline 
\end{tabular}
\caption{$\mathcal{G}$ at zero recoil and $\Sigma_i=\frac{\mathcal{G}_i}{\mathcal{G}_{i-1}}$ on O7.} 
\end{table} 

\end{appendix}
\FloatBarrier
%\appendix
 
\end{document}